\documentclass[12pt,a4paper]{article}

\usepackage[font=small,labelfont=bf,margin=0mm,labelsep=period,tableposition=top]{caption}
\usepackage[a4paper,top=3cm,bottom=3cm,left=2.5cm,right=2.5cm,bindingoffset=0mm]{geometry}

\usepackage{graphicx,ragged2e}
\usepackage{float}
\usepackage{afterpage}
\usepackage{epsfig,cite}
\usepackage{amssymb}
\usepackage{amsmath}
\usepackage{bm}
\usepackage{dsfont}
\usepackage{multirow}
\usepackage{url}
\usepackage{xcolor}
\usepackage{float}
\usepackage{afterpage}
\usepackage{ulem}
\usepackage{url}
\usepackage{multirow,booktabs,multirow}
\newcommand{\be}{\begin{equation}}
\newcommand{\ee}{\end{equation}}
\newcommand{\bea}{\begin{eqnarray}}
\newcommand{\eea}{\end{eqnarray}}
\newcommand{\ei}{\end{itemize}}
\newcommand{\ben}{\begin{enumerate}}
\newcommand{\een}{\end{enumerate}}
\newcommand{\la}{\left\langle}
\newcommand{\ra}{\right\rangle}
\newcommand{\lc}{\left[}
  
  \newcommand{\mr}{\midrule}
  \newcommand{\br}{\bottomrule}
\newcommand{\rc}{\right]}
\newcommand{\lp}{\left(}
\newcommand{\rp}{\right)}

\def\frac#1#2{{{#1}\over {#2}}}
\def\gsim{\mathrel{\rlap{\lower4pt\hbox{\hskip1pt$\sim$}}
    \raise1pt\hbox{$>$}}}       
\def\lsim{\mathrel{\rlap{\lower4pt\hbox{\hskip1pt$\sim$}}
    \raise1pt\hbox{$<$}}}

\newcommand{\rep}{\mathrm{rep}}

\newcommand{\draft}[1]{}
\def\lapprox{\lower .7ex\hbox{$\;\stackrel{\textstyle <}{\sim}\;$}}
\def\gapprox{\lower .7ex\hbox{$\;\stackrel{\textstyle >}{\sim}\;$}}

\numberwithin{equation}{section}
\numberwithin{figure}{section}
\numberwithin{table}{section}

\usepackage{tabularx}
\newcolumntype{C}[1]{>{\centering\arraybackslash}p{#1}}
\begin{document}
\newgeometry{top=1.5cm,bottom=1.5cm,left=2.5cm,right=2.5cm,bindingoffset=0mm}

%\title[Charting Electron Energy Loss Spectroscopy with machine learning]{Charting the low-loss region in Electron Energy Loss Spectroscopy with machine learning}

%\author{}
%\address{}

%\ead{s.conesaboj@tudelft.nl}
%\vspace{10pt}
%\begin{indented}
%\item[]September 2020
%\end{indented}

\begin{flushright}
Nikhef/2020-022\\
\end{flushright}
\vspace{0.3cm}

\begin{center}
  {\Large \bf Charting the low-loss region in Electron Energy \\[0.3cm] Loss Spectroscopy with machine learning}
\vspace{1.4cm}

Laurien I. Roest$^{1,2}$, Sabrya E. van Heijst$^{1}$,
  Louis  Maduro$^{1}$,
  Juan Rojo$^{2,3}$,\\[0.2cm] and Sonia Conesa-Boj$^{1,*}$

\vspace{1.0cm}
 
{\it \small

$^{1}$Kavli Institute of Nanoscience, Delft University of Technology, 2628CJ Delft, The
  Netherlands\\[0.1cm]
$^{2}$Nikhef Theory Group, Science Park 105, 1098 XG Amsterdam, The
  Netherlands \\[0.1cm]$^{3}$Department of Physics and Astronomy, VU,
    1081 HV Amsterdam, The Netherlands

}

\vspace{1.0cm}

{\bf \large Abstract}

\end{center}

Exploiting the information provided by electron energy-loss spectroscopy  (EELS)
requires reliable access to the
low-loss region 
where the zero-loss peak (ZLP) often overwhelms
the contributions associated to inelastic scatterings off the specimen.
Here we deploy machine learning techniques developed in particle physics
to realise a model-independent, multidimensional determination of the ZLP
with a faithful uncertainty estimate.
This novel method 
is then applied to subtract the ZLP for EEL spectra acquired in 
flower-like WS$_2$
nanostructures characterised by  a 2H/3R mixed polytypism.
From the resulting subtracted spectra we
determine the nature and value of the bandgap of polytypic WS$_2$,
finding  $E_{\rm BG} = 1.6_{-0.2}^{+0.3}\,{\rm eV}$ with a clear preference for
an indirect bandgap.
Further, we demonstrate how this method enables us to robustly identify excitonic
transitions down to very small energy losses.
Our approach has been implemented and made available in an
open source {\sc Python} package dubbed {\tt EELSfitter}.

\vspace{0.4cm}
\noindent{\it Keywords:} {\small Transmission Electron Microscopy,
Electron Energy Loss Spectroscopy, Neural Networks, Machine Learning, Transition
Metal Dichalcogenides, Bandgap.}\\

\noindent
{\it $^{*}$corresponding author:} \url{s.conesaboj@tudelft.nl}

\clearpage
\tableofcontents

\section{Introduction}
\label{sec:introduction}

Electron energy-loss spectroscopy (EELS) within the transmission electron microscope (TEM) provides
a wide range of
valuable information on the structural, chemical, and electronic properties of nanoscale materials.
Thanks to recent instrumentation breakthroughs
such as electron monochromators~\cite{Terauchi:2005, Freitag:2005} and aberration correctors~\cite{Haider:1998},
modern EELS analyses can study these properties with highly competitive spatial and spectral resolution.
A particularly important region of EEL spectra is
the low-loss region, defined by electrons that have lost a few tens of eV,
$\Delta E\lsim 50$ eV,
following their inelastic interactions with the sample.
The analysis of this low-loss region makes possible charting the local
electronic properties of nanomaterials~\cite{Geiger:1967}, from the characterisation of
bulk and surface plasmons~\cite{Schaffer:2008}, excitons~\cite{Erni:2005}, 
inter- and intra-band transitions~\cite{Rafferty:1998},
and phonons to the determination of their bandgap and band structure~\cite{Stoger:2008}.

Provided the specimen is electron-transparent, as required for TEM inspection,
the bulk of the incident electron beam will traverse it
either without interacting or restricted to elastic scatterings with the atoms
of the sample's crystalline lattice.
In EEL spectra, these electrons are recorded as a narrow,
high intensity peak centered at energy losses
of $\Delta E\simeq 0$, known as the zero-loss peak (ZLP).
The energy resolution of EELS analyses is ultimately determined by
the electron beam size of the system, often expressed in terms
of the full width at half maximum (FWHM) of the
ZLP~\cite{Egerton:2009}.
In the low-loss region, the contribution from the ZLP
often overwhelms that from the inelastic scatterings arising from
the interactions of the beam electrons with the sample.
Therefore, relevant signals of low-loss phenomena such as excitons,
phonons, and intraband transitions risk becoming drowned
in the ZLP tail~\cite{Abajo:2010}.
An accurate removal of the ZLP
contribution is thus crucial in order to accurately chart and identify the features
of the low-loss region in EEL spectra.

In monochromated EELS, the properties of the ZLP depend on the electron energy dispersion,
the monochromator alignment, and the sample thickness~\cite{Park:2008, Stoger:2008}.
The first two factors arise already in the absence of a specimen (vacuum operation),
while the third is associated
to interactions with the sample such as atomic scatterings,
phonon excitation, and exciton losses.
This implies that EEL measurements in vacuum can be used for calibration purposes
but not to subtract the ZLP from spectra taken on specimens, since their shapes will
in general differ.

Several approaches to ZLP subtraction\cite{Rafferty:2000, Stoger:2008, Egerton:1996} 
have been put forward in the literature.
These are often based on specific model assumptions about the ZLP properties, in particular
concerning its parametric functional dependence on the electron energy loss $\Delta E$,
from Lorentzian~\cite{Dorneich:1998}
and power laws~\cite{Erni:2005} to more general multiple-parameter functions~\cite{Benthem:2001}.
Another approach is based on mirroring the $\Delta E <0$ region of the spectra, assuming
that the $\Delta E>0$ region is fully symmetric~\cite{Lazar:2003}.
More recent studies use integrated software applications for background
subtraction~\cite{Egerton:10.1016/S0304-3991(01)00155-3, Held:2020, Granerod:2018, Fung:2020}.
These various methods are however affected by three main limitations.
Firstly, their reliance on model assumptions such as
the choice of fit function introduces a methodological
bias whose size is difficult to quantify.
Secondly, they lack an estimate of the associated uncertainties, which in turn affects
the reliability of any physical interpretations of the low loss region.
Thirdly, {\it ad hoc} choices such as those of the fitting ranges introduce a significant degree of
arbitrariness in the procedure.

In this study we bypass these limitations by developing a model-independent strategy
to realise a multidimensional determination of the ZLP
with a faithful uncertainty estimate.
Our approach is based on machine learning (ML) techniques
originally developed in high-energy physics to study the
quark and gluon substructure of protons in particle collisions~\cite{Ball:2008by,Ball:2012cx,Ball:2014uwa,Ball:2017nwa}.
It is based on the Monte Carlo replica method to construct a probability
distribution in the space of experimental data and artificial
neural networks as unbiased interpolators to parametrise the ZLP.
The end result is
a faithful sampling of the probability distribution in the ZLP space 
which can be used to subtract its contribution to EEL spectra while
propagating the associated uncertainties.
One can also extrapolate the predictions from this ZLP parametrisation to other TEM
operating conditions beyond those included in the training dataset.

This work is divided into two main parts.
In the first one, we construct a ML model of ZLP spectra acquired
in vacuum, which is able to accommodate an arbitrary number of input
variables corresponding to different operation settings of the TEM.
We demonstrate how this model successfully describes the
input spectra and we assess its extrapolation capabilities for other operation
conditions.
In the second part, we construct a one-dimensional model
of the ZLP as a function of $\Delta E$ from spectra acquired on two different specimens of
tungsten disulfide (WS$_2$) nanoflowers characterised by a 2H/3R mixed polytypism~\cite{SabryaWS2}.
The resulting subtracted spectra are used to determine
the value and nature of the WS$_2$ bandgap in these nanostructures
as well as to map the properties of the associated exciton peaks appearing in the ultra-low
loss region.

This paper is organized as follows.
First of all, in Sect.~\ref{sec:tmdeels}
we review the main features of EELS and present
the WS$_2$ nanostructures that will be used as proof of concept of our approach.
In Sect.~\ref{sec:methodology} we describe the machine learning methodology
adopted to model the ZLP features.
Sects.~\ref{sec:results_vacuum} and~\ref{sec:results_sample} contain
the results of the ZLP parametrisation of spectra acquired
in vacuum and in specimens respectively, which in the latter
case allows us to probe the local electronic properties 
of the WS$_2$ nanoflowers.
Finally in Sect.~\ref{sec:summary} we summarise
and outline possible future developments.
Our results have been obtained with an open-source {\sc Python} code,
dubbed {\tt EELSfitter}, whose installation and usage instructions
are described in Appendix~\ref{sec:installation}.

\section{EELS analyses and TMD nanostructures}
\label{sec:tmdeels}

In this work, we will apply our machine learning method to the study
of the low-loss EELS region of a specific type of WS$_2$ nanostructures presented in~\cite{SabryaWS2},
characterised by a flower-like morphology and a 2H/3R mixed polytypism.
WS$_2$ is a member of the transition metal dichalcogenide (TMD) family, which in turn
belongs to a class of materials known as two-dimensional, van der Waals, or simply layered materials.
These materials are
characterised by the remarkable property of being fully functional down to a single atomic layer.
In order to render the present work self-contained and accessible to a wider audience,
here we review the basic concepts underlying  the EELS technique, and then
present the main features of the WS$_2$ nanoflowers that will be studied in the subsequent sections.

\subsection{EELS and its ZLP in a nutshell}
\label{sec:eels}

\begin{figure}[t]
  \centering
  \includegraphics[width=0.40\textwidth,angle=-90]{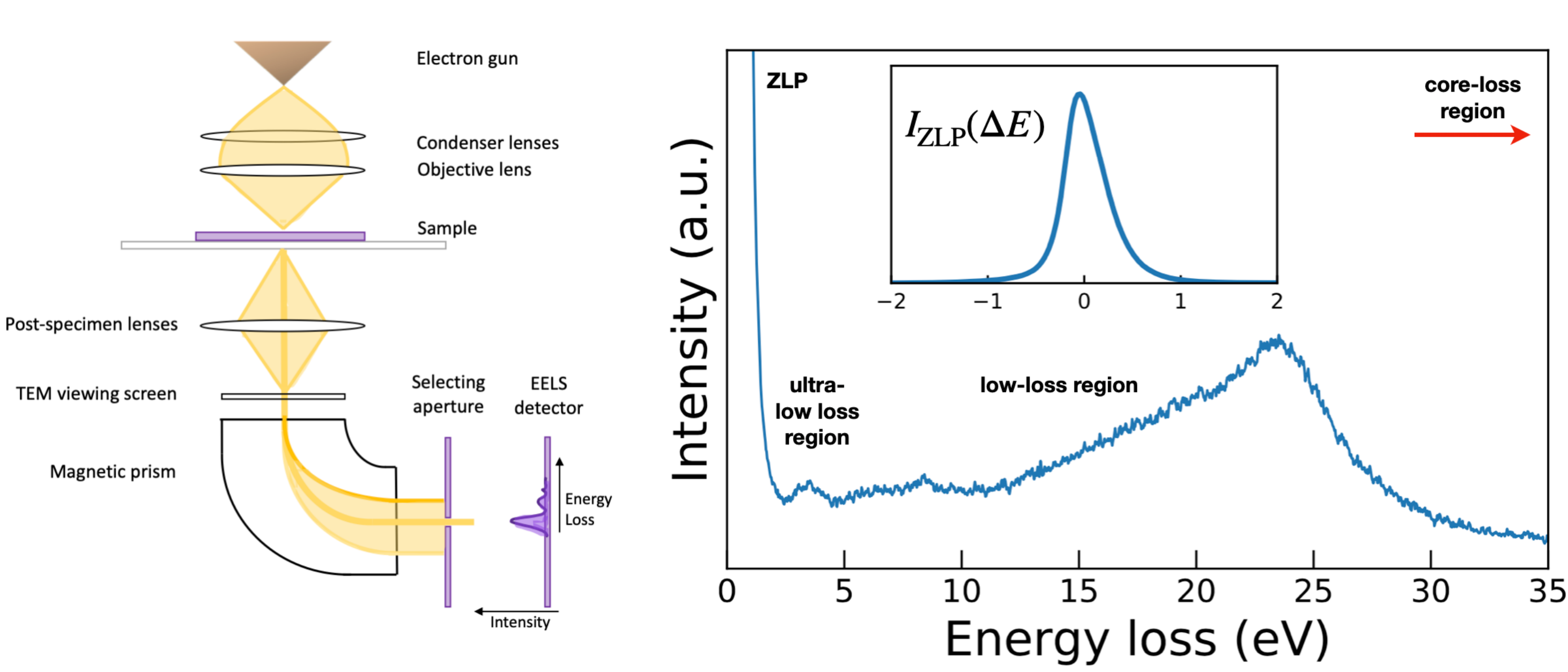}
  \caption{Left: schematic representation of the STEM-EELS setup.
    A magnetic
    prism is used to deflect the electron beam after it has crossed the sample,
   allowing the distribution of energy losses $\Delta E$ to be recorded
    with a spectrometer.
    Right: a representative low-loss EEL spectrum acquired 
    on a WS$_2$ nanoflower~\cite{SabryaWS2} with
    the inset displaying the corresponding ZLP.
  }
  \label{fig:EELS}
\end{figure}

Electron energy loss spectroscopy is a TEM-based method
whereby an electron-transparent sample is illuminated by a 
beam of energetic electrons.
Subsequent to the crossing of
the specimen, the scattered electron beam is focused by a magnetic prism
towards a spectrometer where the distribution of electron energy losses $\Delta E$ can be recorded.
The schematic illustration of a typical EELS setup is shown in the left panel of Fig.~\ref{fig:EELS}.
EEL spectra can be recorded either in the Scanning Transmission Electron Microscopy (STEM) mode
or in the conventional TEM (c-TEM) setup.
Thanks to recent progress in TEM instrumentation and data acquisition, state-of-the-art EELS analyses benefit from
a highly competitive energy (spectral) resolution combined with an unparalleled spatial resolution.

%%%%%%%%%%%%%%%%%%%%%%%%%%%%%%%%%%%%%%%%%%%%%%%
%%%%%%%%%%%%%%%%%%%%%%%%%%%%%%%%%%%%%%%%%%%%%%%%5

EELS spectra can be approximately divided into three main regions.
The first is the zero-loss region, centered around $\Delta E=0$
and containing the contributions from both elastic scatterings
as well as those from electrons that have not interacted with the
sample.
This region is characterised by the strong and narrow ZLP
which dominates over the contribution
from inelastic scatterings.
The second region is the low-loss region, defined for energy losses
$\Delta E \lsim 50$ eV, which contains information
about several important features such as plasmons, excitons, phonons, and
intra-band transitions.
Of particular relevance in this context is the ultra-low loss region, characterised by $\Delta E \simeq$ few eV.
There, the contributions of the ZLP and those from inelastic interactions
become comparable.
The regime for which $\Delta E \gsim 50$ eV is then known as the core-loss region and
provides compositional information
on the materials that constitute the specimen.

The right panel of Fig.~\ref{fig:EELS} displays
a representative EELS spectrum in the region $\Delta E \le 35$ eV, recorded
in one of the WS$_2$ nanoflowers of~\cite{SabryaWS2}.
The inset displays the ZLP, illustrating how nearby $\Delta E\simeq 0$
its size is larger than the contribution from the inelastic scatterings
off the sample by several orders of magnitude.
Carefully disentangling these two contributions 
is essential for the physical interpretation of EEL spectra in the ultra-low-loss region.

The magnitude and shape of the ZLP intensity is known to depend not only on the specific values
of the electron energy loss $\Delta E$, but also on other operation parameters
of the TEM such as the electron beam energy $E_{b}$, the exposure time
$t_{\rm exp}$, the aperture width, and the use of a monochromator.
Since it is not possible to compute the dependence of the ZLP on $\Delta E$
and the other operation parameters from first principles,
reliance on specific models seems to be unavoidable.
This implies that one cannot measure the ZLP for a given operating
condition, for instance a high beam voltage of 200 kV, and expect to reproduce
the ZLP intensity distribution
associated to different conditions, such as a lower beam voltage of 60 kV,
without introducing model assumptions.

Several attempts to describe the ZLP distribution have reported
some success at predicting the main intensity of the peak, 
but in the tails discrepancies are as large as several tens of percent~\cite{Bangert:2003}.
The standard method for background subtraction
is to fit a power law to the tails, however this approach is not suitable in
many circumstances~\cite{Hachtel:2018, Tenailleau:1992, Reed:2002, Bosman:2006}.
Further, even for nominally identical operating conditions, the intensity of the ZLP
will in general vary due to {\it e.g.} external perturbations such as electric or magnetic fields~\cite{Rafferty:2000},
the stability of the microscope and spectrometer electronics~\cite{Kothleitner:2003}, the local
environment (possibly exposed to mechanical, pressure and temperature fluctuations) 
and spectral aberrations~\cite{Egerton:1996}. 
Any robust statistical model for the ZLP should thus account for this irreducible source of uncertainties.

\subsection{TMD materials and WS$_2$ nanoflowers}
\label{sec:tmd}

In this work we will apply our ZLP parametrisation strategy
 to a novel class of recently presented WS$_2$ nanostructures known
as nanoflowers~\cite{SabryaWS2}.
WS$_2$ belongs to the TMD class of layered materials together with {\it e.g.}
MoS$_2$ and WSe$_2$.
TMD materials are of the form MX$_2$, where M is a 
transition metal atom (such as Mo or W) and X a chalcogen atom (such as S, Se, or Te). 
The characteristic crystalline structure of TMDs is such that
one layer of M atoms is sandwiched between two layers of X atoms.

The local electronic structure of TMDs strongly depends on the coordination 
between the transition metal atoms, giving rise to an array of remarkable electronic
and magnetic properties~\cite{Chhowalla:2013}.
Furthermore, the properties of this class of materials vary significantly
with their thickness, for instance MoS$_2$ exhibits an indirect bandgap
in the bulk form which becomes direct at the monolayer level~\cite{Splendiani:2010}.
The tunability of their electronic properties and the associated
potential applications in nano-electronics make TMD materials highly attractive for fundamental research. 

As for other TMD materials, WS$_2$ adopts a layered structure 
by stacking atomic layers of S-W-S in a sandwich-like configuration. 
Although the interaction between adjacent layers is a weak Van der Waals 
force, the dependence of the interlayer interactions on the stacking 
order of WS$_2$ can be significant.
Therefore, modulating the stacking arrangement of WS$_2$ layers (as well
as their relative orientation)
represents a promising handle to tailor the resulting local electronic properties.
WS$_2$ also exhibits a marked thickness dependence of
its properties, with an indirect-to-direct bandgap transition when going
from bulk to bilayer or monolayer form.
The effects of this transition are manifested for example as enhanced
photoluminescence in monolayer WS$_2$, whereas greatly suppressed emission is observed in
the corresponding bulk form~\cite{Zhao2013}.
Further applications of this material include storage of hydrogen 
and lithium for batteries~\cite{Bhandavat:2012}.

%%%%%%%%%%%%%%%%%%%%%%%%%%%%%%%%%%%%%%%%%%%%%%%%%%%%%%%%%%%%%%%%%%%%%%%%%%%%%%%
\begin{figure}[t]
    \centering
    \includegraphics[width=0.99\textwidth]{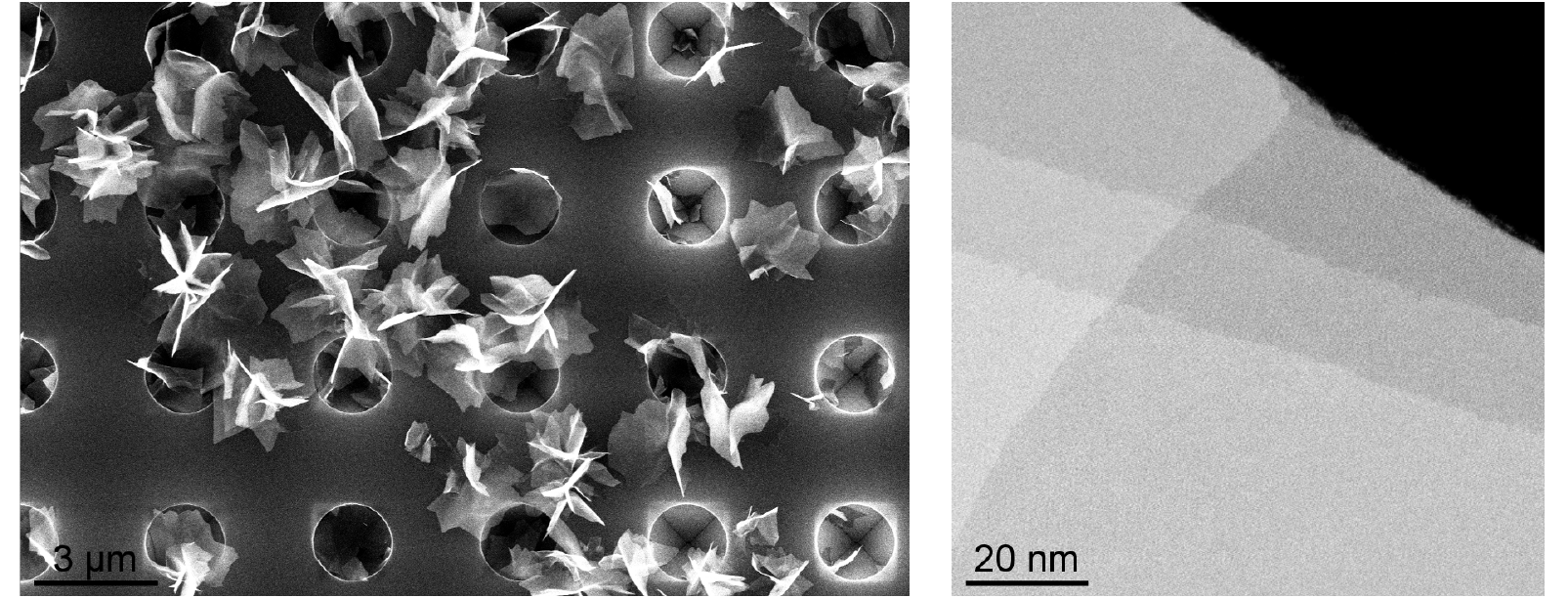}
    \caption{Left: low-magnification TEM image of the WS$_2$ nanoflowers
      grown on top of a holey Si/SiN substrate.
      Right: the magnification of a
      representative petal
      of a nanoflower, where the black region corresponds to the vacuum (no substrate)
      and the difference in contrast indicates terraces of varying thickness.
       }
    \label{fig:nanoflowers}
\end{figure}
%%%%%%%%%%%%%%%%%%%%%%%%%%%%%%%%%%%%%%%%%%%%%%%%%%%%%%%%%%%%%%%%%%%%%%%%%%%%%%%%%%5

A low-magnification TEM image of the WS$_2$ nanoflowers is displayed
in the left panel of Fig.~\ref{fig:nanoflowers}.
These nanostructures are grown directly on top of a holey TEM substrate.
The right panel shows the magnification of a
representative petal
of a nanoflower, where
the difference in contrast indicates terraces of varying thickness.
Note that the black region corresponds to the vacuum, that is, without
substrate underneath.
These WS$_2$ nanoflowers exhibit a wide variety of thicknesses, orientations
and crystalline structures, therefore representing an ideal laboratory to correlate
structural morphology in WS$_2$ with electronic properties at the nanoscale.
Importantly, these nanoflowers are characterised by a mixed crystalline structure,
in particular 2H/3R polytypism.
This implies that different stacking types tend to coexist,  affecting
the interlayer interactions within WS$_2$
and thus modifying the resulting physical properties~\cite{Na:2018}.
One specific consequence of such variations in the stacking patterns is the appearance of
spontaneous electrical polarization, leading to modifications of the 
electronic band structure and thus of the bandgap~\cite{Lee:2016,XIA20171}.

As mentioned above, one of the most interesting properties of  WS$_2$ is
that when the material
is thinned down to a single monolayer its indirect bandgap of
$E_{\rm BG}\simeq 1.4$ eV
switches to a direct bandgap of approximately $E_{\rm BG}\simeq 2.1$ eV.
It has been found that the type and magnitude of the  WS$_2$  bandgap
depends quite sensitively on the crystalline structure and
the number of layers that constitute the material.
In Table~\ref{table:bgvalues} we collect
representative results for the determination of the bandgap energy $E_{\rm BG}$
and its type in WS$_2$, obtained by means of different experimental and theoretical techniques.
 For each reference we indicate separately the bulk results and those
obtained at the monolayer level.
We note that for the latter case there is a fair spread of results in the
value of $E_{\rm BG}$, reflecting the challenges of its accurate determination.

%%%%%%%%%%%%%%%%%%%%%%%%%%%%%%%%%%%%%%%%%%%%%%%%%%%%%%%%%%%%%%%%%%%%%%%%%%%%%%%%%%%%%
\begin{table}[t]
  \small
  \begin{centering}
   \renewcommand{\arraystretch}{1.20}
\begin{tabular}{ccccc}
\br
Reference                       & Thickness & $E_{\rm BG}$ (eV)  & bandgap type  & Technique \\
\mr
{\cite{Braga:2012}} & bulk   & $1.4\pm0.07$            & indirect  & {Gate-voltage dependence}  \\
\mr
\multirow{2}{*}{\cite{Jo:2014}}                 & monolayer  & $2.14 $         & direct  & \multirow{2}{*}{Gate-voltage dependence}        \\
& bulk & $1.40 $    & indirect              \\
\mr

\multirow{2}{*}{\cite{Gusakova:2007}} & monolayer   & $2.03\pm0.03$            & direct  & \multirow{2}{*}{Density Functional Theory}  \\
& bulk & $1.32\pm0.03 $            & indirect     \\
\mr
\multirow{2}{*}{\cite{Kam:1982}}                  & monolayer  & $1.76\pm0.03 $      & direct    & \multirow{2}{*}{Absorption edge coefficient fitting}         \\
& bulk & $1.35 $          & indirect        \\
\mr
\cite{Shi:2013}                &monolayer   & $2.21\pm0.3 $         & direct  & Bethe-Salpeter equation (BSE)        \\                 \br                                         
\end{tabular}
\vspace{0.27cm}
\caption{Representative results for the determination of the bandgap energy $E_{\rm BG}$
  and its type in WS$_2$, obtained by means of different experimental and theoretical techniques.
  For each reference we indicate separately the bulk results and those
  obtained at the monolayer level.}
    \label{table:bgvalues}
    \end{centering}
\end{table}
%%%%%%%%%%%%%%%%%%%%%%%%%%%%%%%%%%%%%%%%%%%%%%%%%%%%%%%%%%%%%%%%%%%%%%%%%%%%%%%%%%%%%%

%%%%%%%%%%%%%%%%%%%%%%%%%%%%%%%%%%%%%%%%%%%%%%%%%%%%%%
\section{A neural network determination of the ZLP}
%%%%%%%%%%%%%%%%%%%%%%%%%%%%%%%%%%%%%%%%%%%%%%%%%%%%%
\label{sec:methodology}

In this section we present our strategy to parametrise and subtract in a model-independent manner
the zero-loss peak that arises in the low-loss region of EEL spectra by means
of machine learning.
As already mentioned, our strategy follows the 
NNPDF approach~\cite{Rojo:2018qdd} originally developed in the context of high-energy physics
for studies of the quark and gluon substructure of the proton~\cite{Gao:2017yyd}.
The NNPDF approach has been successfully applied, among others, to
the determination of the
unpolarised~\cite{DelDebbio:2007ee,Ball:2008by,Ball:2012cx,Ball:2014uwa,Ball:2017nwa}
and polarised~\cite{Nocera:2014gqa} parton distribution functions of protons,  nuclear
parton distributions~\cite{AbdulKhalek:2019mzd,AbdulKhalek:2020yuc}, and the
fragmentation functions of partons into neutral and charged
hadrons~\cite{Bertone:2017tyb,Bertone:2018ecm}.

We note that recently several applications of machine learning
to transmission electron microscopy analyses 
in the context of material science have been
presented, see {\it e.g.}~\cite{Gordon:2020, Zhang:2019, Jany:2017, Ziatdinov:2017,10.1145/2834892.2834896,doi:10.1021/acsnano.7b07504,cite-key}.
Representative examples
include the automated identification
of atomic-level structural information~\cite{10.1145/2834892.2834896},
the extraction of chemical information
and defect classification~\cite{doi:10.1021/acsnano.7b07504},
and spatial resolution enhancement
by means of generative adversarial networks~\cite{cite-key}.
To the best of our knowledge, this is
the first time that neural networks are used as 
 unbiased
 background-removal interpolators
 and combined with Monte Carlo sampling to construct a faithful estimate
of the model uncertainties.

In this section
first of all we discuss the parametrisation of the ZLP in terms of neural networks.
We then review the Monte Carlo replica method used to estimate and propagate the
uncertainties from the input data to physical predictions.
Subsequently, we present our training strategy both in case of vacuum and of sample spectra,
and discuss how one can select the optimal values of the hyper-parameters that appear in the model.

\subsection{ZLP parametrisation}
\label{sec:parametrisation}

To begin with we note that, without any loss of generality, the intensity profile
associated to a generic EEL spectrum may be decomposed as
\be
\label{eq:IeelTot}
I_{\rm EEL}(\Delta E) =I_{\rm ZLP}(\Delta E) + I_{\rm inel}(\Delta E) \, ,
\ee
where $\Delta E$ is the measured electron energy loss; $I_{\rm ZLP}$ is the zero-loss peak
distribution arising both from instrumental origin  and from elastic scatterings; and
$I_{\rm inel}(\Delta E)$ contains the contributions from the
inelastic scatterings off the electrons and atoms in the specimen.
As illustrated by the representative example of Fig.~\ref{fig:EELS}, there are two limits
for which one can cleanly disentangle the two contributions.
First of all, for large enough values of
$\Delta E$ then
$I_{\rm ZLP}$ vanishes and thus $I_{\rm EEL} \to I_{\rm inel}$.
Secondly, in the $\Delta E\simeq 0$ limit all emission can be associated to
 the ZLP such that $I_{\rm EEL}\to  I_{\rm ZLP}$.
In this work we are interested in the ultra-low-loss region, where $I_{\rm ZLP}$ and $I_{\rm inel}$
become of the comparable magnitude.

Our goal is to construct a parametrisation of $I_{\rm ZLP}$ based on artificial
neural networks, which we denote by $I_{\rm ZLP}^{\rm (mod)}$, by means of which one
can extract the inelastic contributions by subtracting the
ZLP background model to the measured intensity spectra,
\be
\label{eq:ZLPseparation}
I_{\rm inel}(\Delta E) \simeq I_{\rm EEL}(\Delta E) - I_{\rm ZLP}^{\rm (mod)}(\Delta E) \, ,
\ee
which enables us to exploit the physical information contained in $I_{\rm inel}$ in
the low-loss region.
Crucially, we aim to faithfully estimate and propagate all the relevant sources of uncertainty associated
both to the input data and to methodological choices.

As discussed in Sect.~\ref{sec:eels}, the ZLP depends both
on the value of the electron energy loss $\Delta E$ as well as on the operation
parameters of the microscope, such as the electron beam energy $E_b$ and the exposure time
$t_{\rm exp}$.
Therefore, we want to construct a multidimensional model which takes all relevant
variables as input.
This means that in general Eq.~(\ref{eq:ZLPseparation}) must be written as
\be
I_{\rm inel}(\Delta E) \simeq I_{\rm EEL}(\Delta E, E_{b},t_{\rm exp}, \ldots) - I_{\rm ZLP}^{\rm (mod)}(\Delta E, E_{b},t_{\rm exp}, \ldots) \, ,
\ee
where we note that the subtracted spectra should depend only on $\Delta E$ but not on the microscope
operation parameters.
Ideally, the ZLP model should be able to accomodate as many input variables as possible.
Here we parametrise $I_{\rm ZLP}^{\rm (mod)}$ by means of
multi-layer feed-forward artificial neural networks~\cite{Forte:2002fg}, that is, we express our ZLP model as
\be
\label{eq:ZLPmodelNN}
I_{\rm ZLP}^{\rm (mod)}(\Delta E, E_{b},t_{\rm exp}, \ldots)  = \xi^{(n_l)}_1(\Delta E, E_{b},t_{\rm exp}, \ldots) \, ,
\ee
where $\xi^{(n_l)}_1$ denotes the activation state of the single neuron in the last
of the $n_l$ layers of the network when the $n_I$ inputs $\{ \Delta E, E_{b},t_{\rm exp}, \ldots \}$
are used.
The weights and thresholds $\{\omega_{ij}^{(l)},\theta_i^{(l)} \}$ of this neural network model are then determined
from the maximization of the model likelihood by means
of supervised learning and non-linear regression from a suitable training dataset.
This type of neural networks benefit from the ability
to parametrise multidimensional input data with arbitrarily
non-linear dependencies: even with a single hidden layer, a neural network
can reproduce arbitrary functional dependencies provided it has a large enough
number of neurons.

\begin{figure}[t]
    \centering
    \includegraphics[width=99mm]{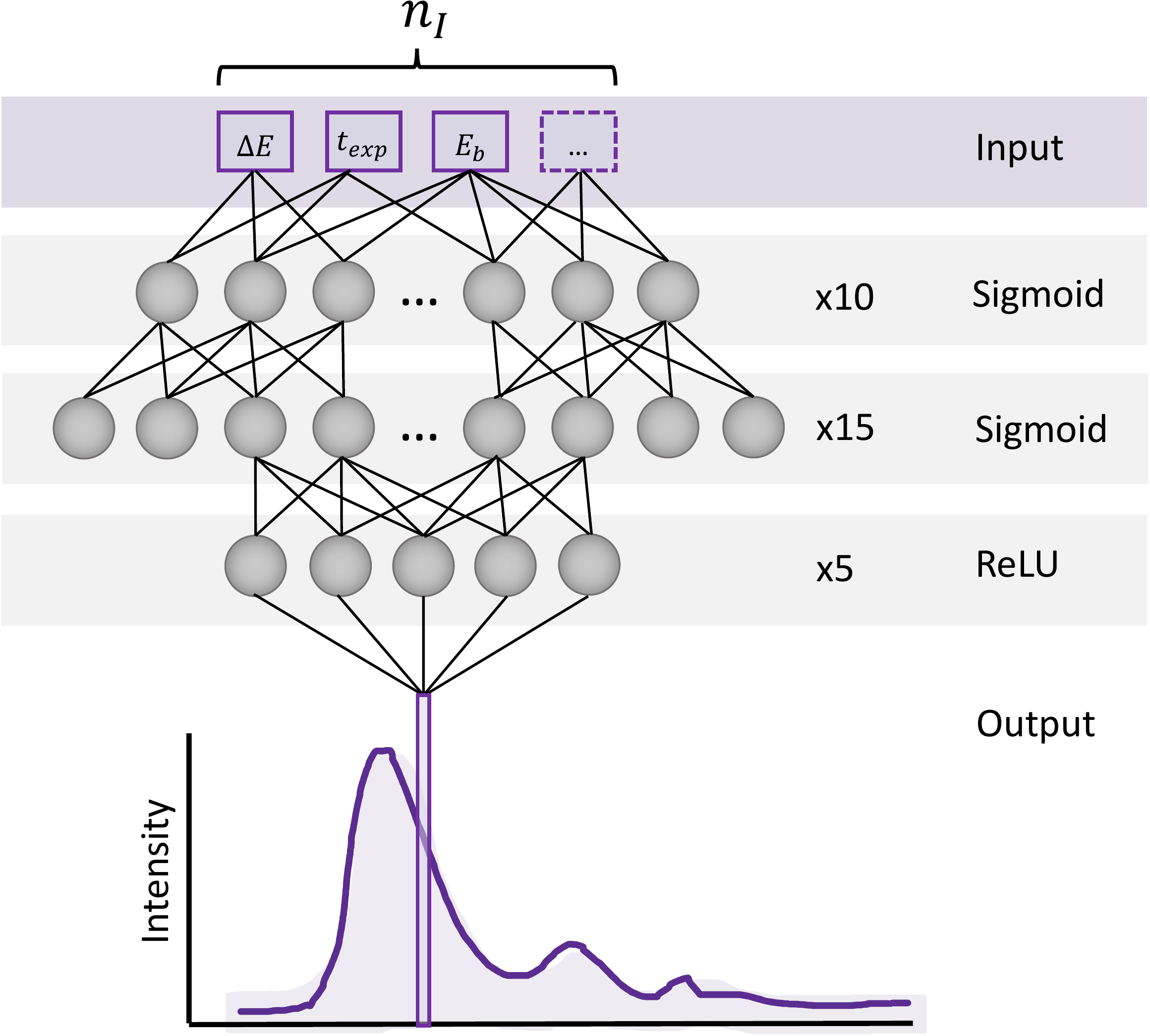}
    \caption{Schematic representation of our neural network model for the ZLP, Eq.~(\ref{eq:ZLPmodelNN}).
      The input is an $n_I$-dimensional array containing $\Delta E$ and other
      operation variables of the microscope such as $E_b$ and $t_{\rm exp}$.
      The output is the predicted value of the intensity of the zero-loss peak
      distribution associated to those specific input variables.
      The architecture is chosen to be $n_I$-10-15-5-1, with sigmoid activation functions
      in all layers except for a ReLU in the output neuron.
    }
    \label{fig:architecture}
\end{figure}

A schematic representation of our model
is displayed in Fig.~\ref{fig:architecture}.
The input is an $n_I$ array containing $\Delta E$ and the rest of
operation variables of the microscope, and
the output is the value of the intensity of the ZLP distribution
associated to those input variables.
We adopt an $n_I$-10-15-5-1 architecture with three hidden layers, for a total
number of 289~(271) free parameters for $n_I=3$~($n_I=1$) to be adjusted by the optimization procedure.
We use a sigmoid activation function for the three hidden layers and a ReLU
for the final one.
The choice of ReLU for the final layer guarantees that our model for the ZLP
is positive-definite, as required by general physical considerations.
We have adopted a redundant architecture  to ensure that the ZLP parametrisation
is sufficiently flexible, and we avoid over-fitting by means of
a suitable regularisation strategy described in Sect.~\ref{sec:training}.
  
%%%%%%%%%%%%%%%%%%%%%%%%%%%%%%%%%%%%%%%%%%%%%
%%%%%%%%%%%%%%%%%%%%%%%%%%%%%%%%%%%%%%%%%%%%%%%%%

\subsection{Uncertainty propagation}
\label{sec:MCreplicas}

We discussed in Sect.~\ref{sec:eels} how
even for EEL spectra taken at nominally identical operation conditions of the microscope,
in general the resulting ZLP intensities will differ.
Further, there exist a large number of different NN configurations, each
representing a different functional form for $I_{\rm ZLP}^{(\rm mod)}$ which provide
an equally valid description of the input data.
To  estimate these uncertainties and propagate them to physical predictions,
we use here the Monte Carlo replica method.
The basic idea  is to exploit the available information
on experimental measurements (central values, uncertainties, and correlations)
to construct a sampling of the probability density in the space of 
the data, which by means of the NN training is then propagated
to a probability density in the space of $I_{\rm ZLP}$ models.

Let us assume that we have $n_{\rm dat}$ independent measurements of the ZLP intensity, for
different or the same values of the input parameters collectively denoted as $\{z_i\}$:
\be
I^{\rm (exp)}_{{\rm ZLP},i}\lp \{ z_i  \}\rp = I^{\rm (exp)}_{{\rm ZLP},i}\lp
\Delta E_i, E_{b,i}, t_{{\rm exp},i},\ldots \rp
\,, \quad i=1,\ldots,n_{\rm dat} \, .
\ee
From these measurements, we can generate a large sample of artificial data points that
will be used as training inputs for the neural nets by means of the   Monte Carlo replica method.
In such approach, one generates  $N_{\rm rep}$ Monte Carlo replicas of the original data points
by means of a multi-Gaussian distribution, with the central values and covariance matrices taken from the input measurements,
\be
\label{eq:MCreplicaGen}
  I_{{\rm ZLP},i}^{{\rm (art)}(k)}  =  I^{\rm (exp)}_{{\rm ZLP},i} + r_i^{({\rm stat},k)}\sigma_i^{\rm (stat)}
  + \sum_{j=1}^{n_{\rm sys}} r_{i,j}^{({\rm sys},k)} \sigma_{i,j}^{\rm (\rm sys)} \,, \quad \forall i
  \,, \quad k=1,\ldots,N_{\rm rep} \,,\,\, \,
  \ee
  where $\sigma_i^{\rm (stat)}$ and $\sigma_{i,j}^{\rm (\rm sys)}$ represent the statistical
  and systematic uncertainties (the latter divided into  $n_{\rm sys}$ fully point-to-point correlated
  sources) and $\{r_i^{(k)}\}$ are Gaussianly distributed random numbers.
  The values of $\{r_i^{(k)}\}$ are
  generated with a suitable correlation pattern to ensure
  that averages over the set of Monte Carlo
  replicas reproduce the original experimental covariance matrix, namely
  \be
  \la  \lp I_{{\rm ZLP},i}^{{\rm (art)}(k)} - \la I_{{\rm ZLP},i}^{{\rm (art)}}\ra_{\rm rep}\rp
  \lp I_{{\rm ZLP},j}^{{\rm (art)}(k)} - \la I_{{\rm ZLP},j}^{{\rm (art)}}\ra_{\rm rep}\rp\ra_{\rm rep}
  \label{eq:expcovariance} = {\rm cov}^{(\rm exp)}\lp I_{{\rm ZLP},i},I_{{\rm ZLP},j}\rp  \, ,
  \ee
  where averages are evaluated over the $N_{\rm rep}$ replicas that compose the sample.
We thus note that each $k$-th replica contains 
as many data points as the original set.

In our case, the information on experimental correlations is not accessible and
thus we assume that there is a single source of point-by-point uncorrelated systematic
uncertainty, denoted as $\sigma_i^{\rm (exp)}$, which is estimated as follows.
The input measurements will be composed in general on subsets of EEL
spectra taken with identical operation conditions.
Assume that for a specific set of operation conditions we have $N_{\rm sp}$ of such spectra.
Since the values of $\Delta E$ will be different in each case, first of all
we uniformise a common binning in $\Delta E$ with $n_{\rm dat}$ entries.
Then we evaluate the total experimental uncertainty in one of these bins as
\be
\label{eq:sigmaiexp}
\sigma_i^{\rm (exp)} = \lp \frac{1}{N_{\rm sp}-1} \sum_{l=1}^{N_{\rm sp}}
\lp I_{{\rm ZLP},i}^{ ({\rm exp}),l}  - \la I_{{\rm ZLP},i}^{ ({\rm exp})}\ra_{N_{\rm sp}} \rp \rp^{1/2} \, ,\,
i=1,\ldots, n_{\rm dat} \, ,
\ee
that is, as the standard deviation over the $N_{\rm sp}$ spectra.
This uncertainty is separately evaluated for each set of microscope operation conditions
for which data available.
In the absence of correlations, Eqns.~(\ref{eq:MCreplicaGen}) and~(\ref{eq:expcovariance})  simplify to
\be
 I_{{\rm ZLP},i}^{{\rm (art)}(k)}  =  I^{\rm (exp)}_{{\rm ZLP},i} + r_i^{({\rm tot},k)}\sigma_i^{\rm (exp)}
 \,, \quad \forall i
  \,, \quad k=1,\ldots,N_{\rm rep} \,.\,\, \,
\ee
and
  \bea
  \la  \lp I_{{\rm ZLP},i}^{{\rm (art)}(k)} - \la I_{{\rm ZLP},i}^{{\rm (art)}}\ra_{\rm rep}\rp
  \lp I_{{\rm ZLP},j}^{{\rm (art)}(k)} - \la I_{{\rm ZLP},j}^{{\rm (art)}}\ra_{\rep}\rp\ra_{\rm rep} =
  \sigma_i^{\rm (exp)}\sigma_j^{\rm (exp)}\delta_{ij} \, ,
  \eea
  since the experimental covariance matrix is now diagonal.
  Should in the future correlations became available, it would be straightforward to extend
  our model to that case.

\begin{figure}[t]
    \centering
    \includegraphics[width=0.99\textwidth]{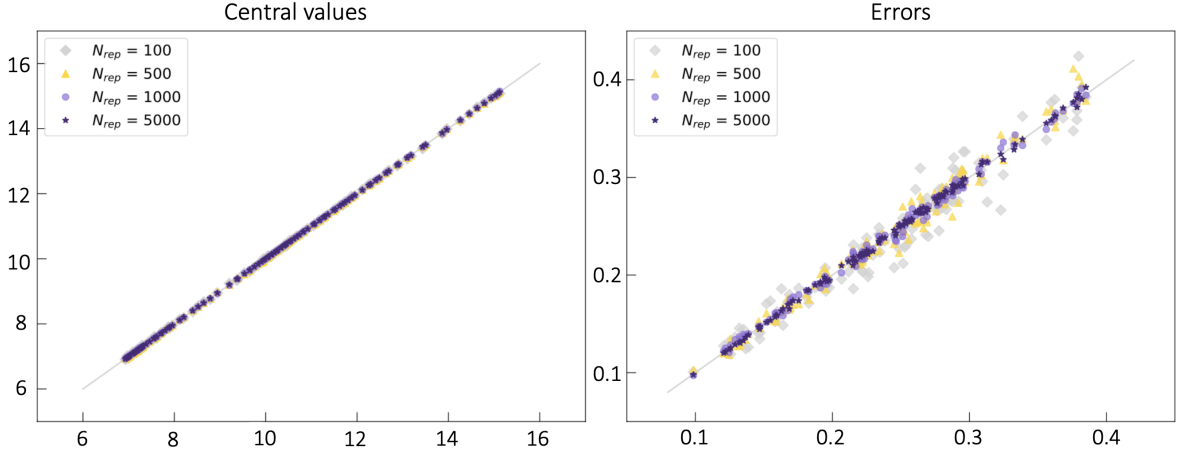}
    \caption{Comparison between the original experimental central values
      $I_{\rm ZLP,i}^{\rm (exp)}$ (left) and the corresponding 
      uncertainties $\sigma_i^{(\rm exp)}$ (right panel) with the results of averaging over
      a sample of $N_{\rm rep}$ Monte Carlo replicas generated by means of
      Eq.~(\ref{eq:MCreplicaGen}), for different values of
      $N_{\rm rep}$.
      }
    \label{fig:MC}
\end{figure}

The value of the number of generated MC replicas, $N_{\rm rep}$, should be chosen such that the set of replicas 
 accurately reproduces the probability distribution of the original training data.
To verify that this is the case,
Fig.~\ref{fig:MC} displays a comparison between the original experimental central values
$I_{{\rm ZLP},i}^{\rm (exp)}$  and the corresponding 
total uncertainties $\sigma_i^{(\rm exp)}$ with the results of averaging over
a sample of $N_{\rm rep}$ Monte Carlo replicas generated by means of
Eq.~(\ref{eq:MCreplicaGen}) for different number of replicas.
We find that $N_{\rm rep}=500$ is a value that ensures that both
the central values and uncertainties are reasonably well reproduced,
and we adopt it in what follows.

%%%%%%%%%%%%%%%%%%%%%%%%%%%%%%%%%%%%%%%%%%%%%%%
%%%%%%%%%%%%%%%%%%%%%%%%%%%%%%%%%%%%%%%%%%%%%%%%5

\subsection{Training strategy}
\label{sec:training}

The training of the neural network model for the ZLP peak differs between
the cases of EEL spectra taken on vacuum, where by construction $I_{\rm EEL}(\Delta E) =I_{\rm ZLP}^{\rm (mod)}(\Delta E)$,
and for spectra taken on specimens\footnote{Actually, EEL spectra taken in the vacuum but close enough
  to the sample might still receive inelastic contributions from the specimen. In this work,
  when we use vacuum spectra, we consider exclusively those acquired reasonably far from the surfaces
of the analysed nanostructures.}.
In the latter case, as indicated by Eq.~(\ref{eq:ZLPseparation}), in order to avoid
biasing the results it is
important to ensure that the model is trained only on the region of the spectra
where the ZLP dominates over the inelastic scatterings.
We now describe the training strategy that is adopted for these two cases.

\paragraph{Training on vacuum spectra.}
For each of the $N_{\rm rep}$ generated Monte Carlo replicas, we train an independent
neural network as described in Sect.~\ref{sec:parametrisation}.
The parameters of the neural network $ \{\theta^{(k)}\}$ (its weights and thresholds)
are determined from the minimisation of a figure of merit (the cost function of the model)
defined as
\begin{equation}
  \label{eq:chi2}
\begin{centering}
  E^{(k)}\lp \{\theta^{(k)}\}\rp = \frac{1}{n_{\rm dat}}\sum_{i=1}^{n_{\rm dat}}\left(\frac{ I_{{\rm ZLP},i}^{{\rm (art)}(k)} -
  I_{{\rm ZLP},i}^{{\rm (mod)}}\lp \{\theta^{(k)}\}\rp }{\sigma_i^{(\rm exp)}}\right)^2, 
\end{centering}
\end{equation}
which is the $\chi^2$ per data point obtained by comparing the $k$-th replica for the ZLP
intensity with the corresponding model prediction for the values
$\{\theta^{(k)}\}$ of its weights and thresholds.
In order to speed up the neural network training process, prior to the optimisation
all inputs and outputs are scaled to lie between $[0.1, 0.9]$ before
being fed to the network.
This preprocessing facilitates that
 the neuron activation states will typically
lie close to the linear region of the sigmoid activation function.

The contribution to the figure of merit from the input experimental data, Eq.~(\ref{eq:chi2}),
needs in general to be complemented with that of theoretical constraints on the model.
For instance, when determining nuclear parton distributions~\cite{AbdulKhalek:2020yuc}, one needs to
extend Eq.~(\ref{eq:chi2}) with Lagrange multipliers to ensure that both the $A=1$ proton boundary
condition and the cross-section positivity are satisfied.
In the case at hand, our model for the ZLP should implement the property that $I_{\rm ZLP}(\Delta E)\to 0$
when $|\Delta E| \to \infty$, since far from $\Delta E\simeq 0$ the contribution from elastic scatterings
and instrumental broadening is completely negligible.
In order to implement this constraint, we add $n_{\rm pd}$ pseudo-data points to the training dataset and modify
the figure of merit Eq.~(\ref{eq:chi2}) as follows
\be
\label{eq:chi2modified}
E^{(k)}\lp \{\theta^{(k)}\}\rp \to E^{(k)}\lp \{\theta^{(k)}\}\rp +
\lambda \sum_{i'=1}^{n_{\rm pd}}\left(
I_{{\rm ZLP},i'}^{{\rm (mod)}}\lp \{\theta^{(k)}\}\rp \right)^2, 
\ee
where $\lambda$ is a Lagrange multiplier whose value is tuned to ensure that the $I_{\rm ZLP}(\Delta E)\to 0$
condition
is satisfied without affecting the description of the training dataset.
The pseudo-data is chosen to lie in the region $[\Delta E_{\rm pd}^{\rm (min)},
  \Delta E_{\rm pd}^{\rm (max)}]$ (and symmetrically for energy gains).

The value of $\Delta E_{\rm pd}^{\rm (min)}$
can be determined automatically by evaluating the ratio $\mathcal{R}_{\rm sig}$ between the central
experimental intensity and the total uncertainty in each data point,
\be
\label{eq:pdlocation}
\mathcal{R}_{\rm sig}(\Delta E_i)\equiv \frac{I_{{\rm ZLP}}^{(\rm exp)}(\Delta E_i)}{\sigma^{(\rm exp)}(\Delta E_i)} \, ,
\ee
which corresponds to the statistical significance for the $i$-th bin of $\Delta E$ to differ from the null hypothesis
(zero intensity) taking into account the experimental uncertainties.
For sufficiently large energy losses one finds that $\mathcal{R}_{\rm sig}(\Delta E)\lsim 1$,
indicating that one would be essentially fitting statistical noise.
In order to avoid such a situation and only fit data that is different from zero within errors, we
determine $\Delta E_{\rm pd}^{\rm (min)}$ from the condition $\mathcal{R}_{\rm sig}\simeq 1$.
We then maintain the training data in the region $\Delta E \le \Delta E_{\rm pd}^{\rm (min)}$ and the pseudo-data
points are added for $[ \Delta E_{\rm pd}^{\rm (min)}, \Delta E_{\rm pd}^{\rm (max)}]$. 
The value of $\Delta E_{\rm pd}^{\rm (max)}$ can be chosen arbitrarily and can be as large as necessary
to ensure that $I_{\rm ZLP}(\Delta E)\to 0$ as $|\Delta E| \to \infty$.

We note that another important physical condition on the ZLP model, namely its positivity
(since in EEL spectra the intensity is just a measure of the number of counts in the
detector for a given value of the energy loss), is automatically satisfied given that
we adopt a ReLU activation function for the last layer.

In this work we adopt the {\tt TensorFlow} library~\cite{Abadi:2016kic} to assemble
the architecture illustrated in  Fig.~\ref{fig:architecture}.
Before training, all weights and biases are initialized in a non-deterministic order
by the built-in global variable initializer. 
The optimisation of the figure of merit Eq.~(\ref{eq:chi2modified}) is carried
out by means of stochastic gradient descent (SGD) combined with backpropagation, specifically
by means of the Adam minimiser.
The hyper-parameters of the optimisation algorithm such as the learning rate
have been adjusted to ensure proper learning is reached in the shortest amount
of time possible.

Given that we have a extremely flexible parametrisation, one should be careful
to avoid overlearning the input data.
Here over-fitting is avoided by means of the following cross-validation stopping criterion.
We separate the input data into training and validation subsets, with a 80\%/20\% splitting
which varies randomly for each Monte Carlo replica.
We then run the optimiser for a very large number of iterations and store both
the state of the network and the value
of the figure of merit Eq.~(\ref{eq:chi2}) restricted to the validation
dataset, $E^{(k)}_{\rm val}$ (which is not used for the training).
The optimal stopping point is then determined {\it a posteriori} for each replica
as the specific network configuration that leads to the deepest minimum of $E^{(k)}_{\rm val}$.
The number of epochs should be chosen high enough to reach the optimal stopping point for each replica.
In this work we find that $40k$ epochs are sufficient to be able to identify these optimal stopping points.
This corresponds to a serial running time of 
$t\simeq 60$ seconds per replica when running the optimization on a  single CPU for 500 datapoints.

Once the training of the $N_{\rm rep}$ neural network models for the ZLP has been carried out,
we gauge the overal fit quality of the model by computing the
$\chi^2$ defined as
\begin{equation}
  \label{eq:chi2_final}
\begin{centering}
  \chi^2 = \frac{1}{n_{\rm dat}}\sum_{i=1}^{n_{\rm dat}}\left(\frac{ I_{{\rm ZLP},i}^{{\rm (exp)}} -
 \la I_{{\rm ZLP},i}^{{\rm (mod)}}\ra_{\rm rep} }{\sigma_i^{(\rm exp)}}\right)^2, 
\end{centering}
\end{equation}
which is the analog of Eq.~(\ref{eq:chi2_final}) now comparing the average model prediction
to the original experimental data values.
A value $\chi^2 \simeq 1$ indicates that a satisfactory description of the experimental data,
within the corresponding uncertainties, has been achieved.
Note that in realistic scenarios $\chi^2$ can deviate from unity, for instance when
some source of correlation between the experimental uncertainties has been neglected, or on the contrary
when the total experimental error is being underestimated.

\paragraph{Training on sample spectra.}

The training strategy for the case of EEL spectra acquired on specimens (rather than on vacuum) must be adjusted
to account for the fact that the input data set, Eq.~(\ref{eq:IeelTot}), receives contributions
both from the ZLP and from inelastic scatterings.
To avoid biasing the ZLP model, only the former contributions should be
included in the training dataset.

\begin{figure}[t]
    \centering
    \includegraphics[width=0.89\textwidth]{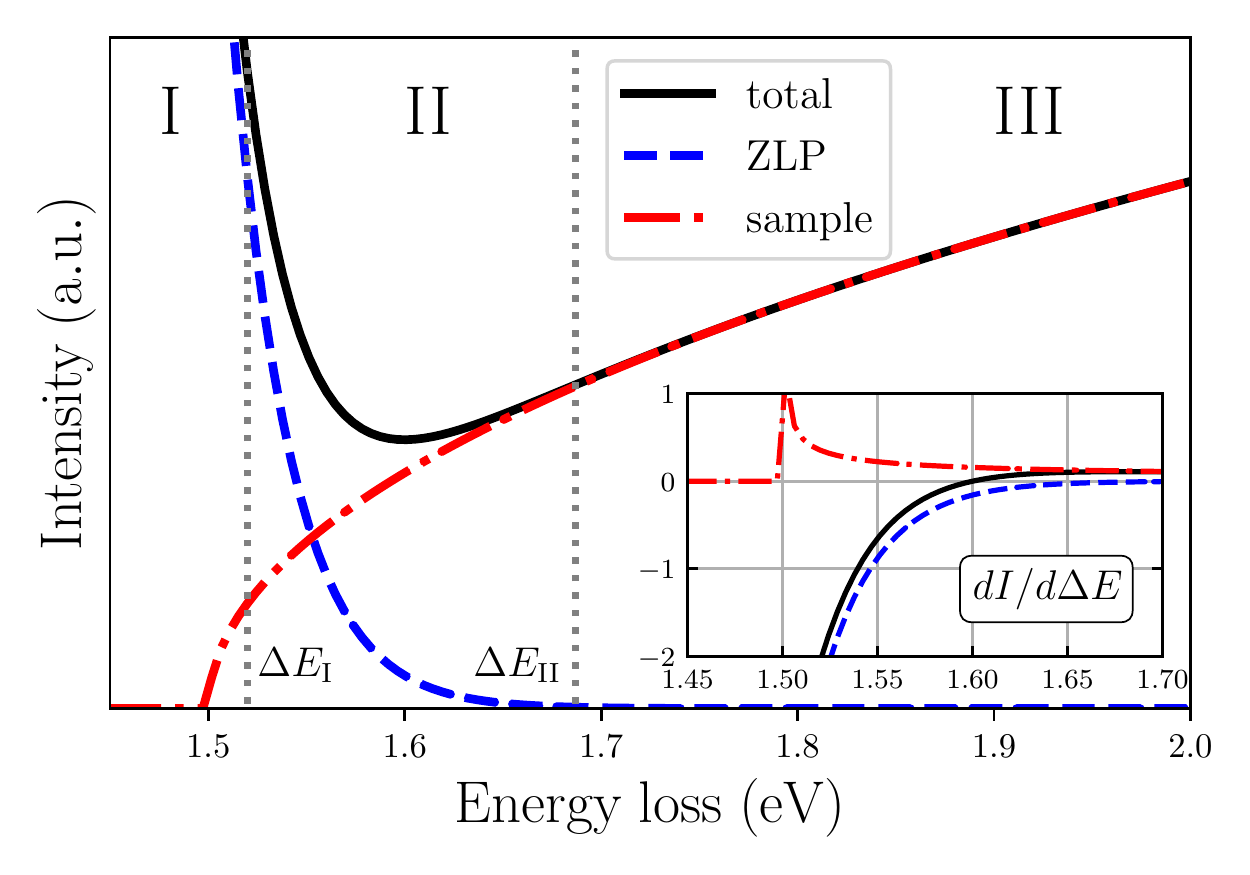}
    \caption{A toy model for the EEL spectrum and its
      derivative (in the inset).
      We display the separate contributions from $I_{\rm ZLP}$
      and $I_{\rm inel}$ as well as their sum (total).
      We indicate the two regions used for the model training ($\rm I$ and $\rm III$),
      while as discussed in the text the neural network predictions
      are
      extrapolated to region $\rm II$, defined by $\Delta E_{\rm I} \le \Delta E \le \Delta E_{\rm II}$.
    }
    \label{fig:EELS_toy}
\end{figure}

We can illustrate the situation at hand with the help of a simple toy model for the low-loss
region of the EEL spectra, represented in
Fig.~\ref{fig:EELS_toy}.
Let us assume for illustration purposes that the ZLP is described by a Gaussian distribution,
\be
I_{\rm ZLP}(\Delta E) \propto  \exp \lp -\frac{\Delta E^2}{\sigma_{\rm ZLP}^2}\rp \, ,
\ee
with a standard deviation of $\sigma_{\rm ZLP}=0.3$ eV,
and that the contribution from the
inelastic scatterings arising from the sample can be approximated in the low-loss
region by
\be
I_{\rm inel}(\Delta E)\propto \lp \Delta E - E_{\rm BG}\rp^b \, ,
\ee
with $E_{\rm BG}=1.5$ eV and $b=1/2$.
The motivation for the latter
choice will be spelled out in Sect.~\ref{sec:results_sample}.
We display the separate contributions from $I_{\rm ZLP}$
and $I_{\rm inel}$, as well as their sum, 
with the inset showing the values of the corresponding derivatives, $dI/d\Delta E$.

%%%%%%%%%%%%%%%%%%%%%%%%%%%%%%%%%%%%%%%%%%%%%
%%%%%%%%%%%%%%%%%%%%%%%%%%%%%%%%%%%%%%%%%%%%%%%%%

While simple, the toy model of Fig.~\ref{fig:EELS_toy} is actually general enough so that one can draw
a number of useful considerations concerning the relation between $I_{\rm ZLP}$ and $I_{\rm inel}$
that will apply also in realistic spectra:

\begin{itemize}

\item The ZLP intensity, $I_{\rm ZLP}(\Delta E)$, is a monotonically decreasing function
  and thus its derivative is always negative.

\item  The first local minimum of the total intensity, $dI_{\rm EEL}/d\Delta E|_{\Delta E_{\rm min}}=0$, corresponds
  to a value of $\Delta E$ for which the contribution from the inelastic emissions is already
  sizable.

\item The value of $\Delta E$ for which $I_{\rm inel}$ starts to contribute to the total spectrum
  corresponds to the position where the derivatives of the in-sample and in-vacuum intensities
  start to differ.
  We note that a direct comparison between the overal magnitude of the sample and vacuum ZLP
  spectra is in general not possible, as explained in Sect.~\ref{sec:eels}. 
\end{itemize}

These considerations suggest that when training the ML model on EEL spectra recorded on samples,
the following categorisation should de adopted:

\begin{enumerate}

\item For energy losses $\Delta E \le \Delta E_{\rm I}$ (region $\rm I$),
  the model training  proceeds in exactly the same way as for the vacuum case
  via the minimisation of Eq.~(\ref{eq:chi2}).

\item  
  For $\Delta E \ge \Delta E_{\rm II}$ (region $\rm III$), we use instead Eq.~(\ref{eq:chi2modified})
  without the contribution from the input data, since for such values
  of $\Delta E$ one has that $I_{\rm inel}\gg I_{\rm ZLP}$.
  In other words, the only information that the region $\rm III$ provides
  on the model is the one arising from the implementation
  of the constraint that $I_{\rm ZLP}(\Delta E\to \infty)\to 0$.

\item The EELS measurements  in region $\rm II$, defined by  $\Delta E_{\rm I} \le \Delta E \le \Delta E_{\rm II}$,
  are excluded from the training dataset, given that in this region the contribution to $I_{\rm EEL}$
  coming from $I_{\rm inel}$ is significant.
  There the model predictions are obtained from an interpolation
  of the associated predictions obtained in the regions $\rm I$ and $\rm III$.

\end{enumerate}

The categorisation introduced in Fig.~\ref{fig:EELS_toy} relies on
two hyper-parameters of the model, $\Delta E_{\rm I}$ and
$\Delta E_{\rm II}$, which need to be specified before the training takes place.
They should satisfy $\Delta E_{\rm I} \le \Delta E_{\rm min}$ and $\Delta E_{\rm II} \ge \Delta E_{\rm min}$,
with $\Delta E_{\rm min}$ being the position of the first local minimum of $I_{\rm EEL}$.
As indicated by the toy spectra of Fig.~\ref{fig:EELS_toy}, a suitable value for $\Delta E_{\rm I}$
would be somewhat above the onset of the inelastic contributions, to maximise
the amount of training data while ensuring that $I_{\rm EEL}$ is still dominated
by $I_{\rm ZLP}$.

The optimal value of $\Delta E_{\rm I}$  can be determined as follows.
We evaluate the ratio
between the derivative of the intensity distribution acquired on the specimen over the
same quantity recorded in vacuum,
\be
\label{eq:rder}
\mathcal{R}^{(j)}_{\rm der}(\Delta E) \equiv
\la
\frac{
  dI_{\rm EEL}^{({\rm exp})(j)}(\Delta E)/ d\Delta E
}{
  dI_{\rm EEL}^{({\rm exp})(j')}(\Delta E) /d\Delta E
} \ra_{N_{\rm sp}' } \, ,
\ee
where $j'$ labels one of the $N_{\rm sp}'$ vacuum spectra and the average is taken
over all available values of $j'$.
This ratio allows one to identify a suitable value of $\Delta E_{\rm I}$ by establishing
for which energy losses the shape (rather than the absolute value) of the intensity distributions 
recorded on the specimen starts to differ significantly from their vacuum counterparts.
A sensible choice of $\Delta E_{\rm I}$ could for instance be given by
$\mathcal{R}_{\rm der}(\Delta E_{\rm I}) \simeq 0.8$, for which derivatives differ
at the 20\% level.
Note also that the leftmost value of the energy loss satisfying
$\mathcal{R}_{\rm der}(\Delta E)=0$ in Eq.~(\ref{eq:rder}) corresponds to the position of the first
local minimum.

Concerning the choice of the second hyper-parameter $\Delta E_{\rm II}$, following the discussion
above one can identify $\Delta E_{\rm II}=\Delta E_{\rm pd}^{\rm (min)}$, which is determined by requiring  that
Eq.~(\ref{eq:pdlocation}) satisfies $\mathcal{R}_{\rm sig}(\Delta E_i)\lsim 1$ and thus correspond
to the value of $\Delta E$ where statistical uncertainties drown the signal intensity.

\section{ZLP parametrisation from vacuum spectra}
\label{sec:results_vacuum}

We now move to discuss the application of the strategy presented in the previous
section to the parametrisation of ZLP spectra acquired in vacuum.
Applying our model to this case has a two-fold motivation.
First of all, we aim to demonstrate that the model is sufficiently flexible
to effectively reproduce the
input EELS measurements for a range of variations of the operation parameters of the microscope.
Second, it allows one to provide a calibrated prediction
useful for the case of the in-sample measurements.
Such calibration is necessary since, as explained in Sect.~\ref{sec:training}, some of the model
hyper-parameters are determined by comparing intensity shape profiles
between spectra taken in vacuum and in sample.

In this section, first of all we present the input dataset and motivate the choice
of training settings and model hyperparameters.
Then we validate the model training by assessing the fit quality.
Lastly, we study the dependence of the model output in its various input
variables, extrapolate  its predictions to new operation
conditions, and study the dependence of the model uncertainties upon
restricting the training dataset.

\subsection{Training settings}

In Table~\ref{table:vacuumdata} we collect the main properties of the EELS spectra
acquired in vacuum to train the neural
network model.  For each set of spectra, we indicate the exposure time $t_{\rm exp}$, the beam energy
$E_b$, the number of spectra $N_{\rm sp}$ recorded for these operation conditions, the number $n_{\rm dat}$ of
bins in each spectrum, the range in electron energy loss $\Delta E$,
and the average full width at half maximum (FWHM)
evaluated over the $N_{\rm sp}$ spectra with the corresponding standard deviation.
The spectra  listed on Table~\ref{table:vacuumdata}
were acquired with a ARM200F Mono-JEOL microscope equipped
with a GIF continuum spectrometer, see also Methods.
We point out that since here
we are interested in the low-loss region, $\Delta E_{\rm max}$ does not need
to be too large, and anyway the asymptotic $\Delta E$ behaviour of the model is fixed
by the constraint implemented by Eq.~(\ref{eq:chi2modified}).

%%%%%%%%%%%%%%%%%%%%%%%%%%%%%%%%%%%%%%%%%%%%%%%%%%%%%%%%%%%%%%%%%%%%%%%%%%%%%%%%%%%%%%%%%%%%%
%%%%%%%%%%%%%%%%%%%%%%%%%%%%%%%%%%%%%%%%%%%%%%%%%%%%%%%%%%%%%%%%%%%%%%%%%%%%%%%%%%%%%%%%%%%%%
\begin{table}[t]
  \begin{center}
            \renewcommand{\arraystretch}{1.50}
  \begin{tabular}{@{}ccccccccc}
\br
Set & $t_{\rm exp}$ {(}ms{)} & $E_{\rm b}$ {(}keV{)} & $N_{\rm sp}$ & $n_{\rm dat}$ & $\Delta E_{\rm min}$~(eV)  & $\Delta E_{\rm max}$~(eV)  & FWHM~(meV)  \\ 
\mr
1        & 100                 & 200                  & 15          & 2048               & -0.96              & 8.51     & $47\pm7 $         \\
2        & 100                 & 60                   & 7           & 2048               & -0.54              & 5.59    & 
$ 50 \pm 4$         \\
3        & 10                  & 200                  & 6          & 2048               & -0.75              & 5.18      & 
$ 26 \pm 3$         \\
4        & 10                  & 60                   & 6           & 2048               & -0.40              & 4.78       & 
$ 34\pm 2$         \\ 
\br
  \end{tabular}
    \end{center}
  \caption{\small Summary of the main properties of the EELS spectra acquired in vacuum to train the neural
    network model.  For each set of spectra, we indicate the exposure time $t_{\rm exp}$, the beam energy
    $E_b$, the number of spectra $N_{\rm sp}$ recorded for these operation conditions, the number $n_{\rm dat}$ of
    bins in each spectrum, the range in electron energy loss $\Delta E$,
    and the average FWHM evaluated over the $N_{\rm sp}$ spectra with the corresponding standard deviation
  }
   \label{table:vacuumdata}
\end{table}
%%%%%%%%%%%%%%%%%%%%%%%%%%%%%%%%%%%%%%%%%%%%%%%%%%%%%%%%%%%%%%%%%%%%%%%%%%%%%%%%%%%%%%%%%%%%%%%%%5
%%%%%%%%%%%%%%%%%%%%%%%%%%%%%%%%%%%%%%%%%%%%%%%%%%%%%%%%%%%%%%%%%%%%%%%%%%%%%%%%%%%%%%%%%%%%%

The energy resolution of these spectra, quantified by the average value of their FWHM, ranges
from 26 meV to 50 meV depending on the specific operation conditions of the microscope,
with an standard deviation between 2 and 7 meV.
The value of the FWHM varies only mildly with the value of the beam energy $E_b$
but grows rapidly for spectra collected with larger exposure times $t_{\rm exp}$.
A total of almost $7\times 10^4$ independent measurements will be used for the ZLP model
training on the vacuum spectra.
As will be highlighted in Sects.~\ref{eq:depdeltae} and~\ref{eq:depebeam}, one of the advantages of our ZLP model is that it can extrapolate its predictions
to other operation conditions beyond the specific ones used for the training
and listed in Table~\ref{table:vacuumdata}.

Following the strategy presented in Sect.~\ref{sec:methodology}, first of all we combine the $N_{\rm sp}$ spectra
corresponding to each of the four sets of operation conditions and determine the statistical uncertainty
associated to each energy loss bin by means of Eq.~(\ref{eq:sigmaiexp}).
For each of the training sets, we need to determine the value
of $\Delta E_{\rm pd}^{\rm (min)}~(=\Delta E_{\rm II})$
that defines the range for which we add the  pseudo-data
that imposes the correct $\Delta E \to \infty$ limit of the model.
This value is fixed by the condition that
ratio between the central experimental value of the 
EELS intensity and its corresponding uncertainty,
Eq.~(\ref{eq:pdlocation}), satisfies $\mathcal{R}_{\rm sig}\simeq 1$.
      
%%%%%%%%%%%%%%%%%%%%%%%%%%%%%%%%%%%%%%%%%%%%%%%%%%%%%%%%%%%%%%%%%%%%%%%%%%%%%%%%%%%%%%%%%%%%%%%%%%%%%%%%%%%%
\begin{figure}[t]
  \centering
  \includegraphics[width=125mm]{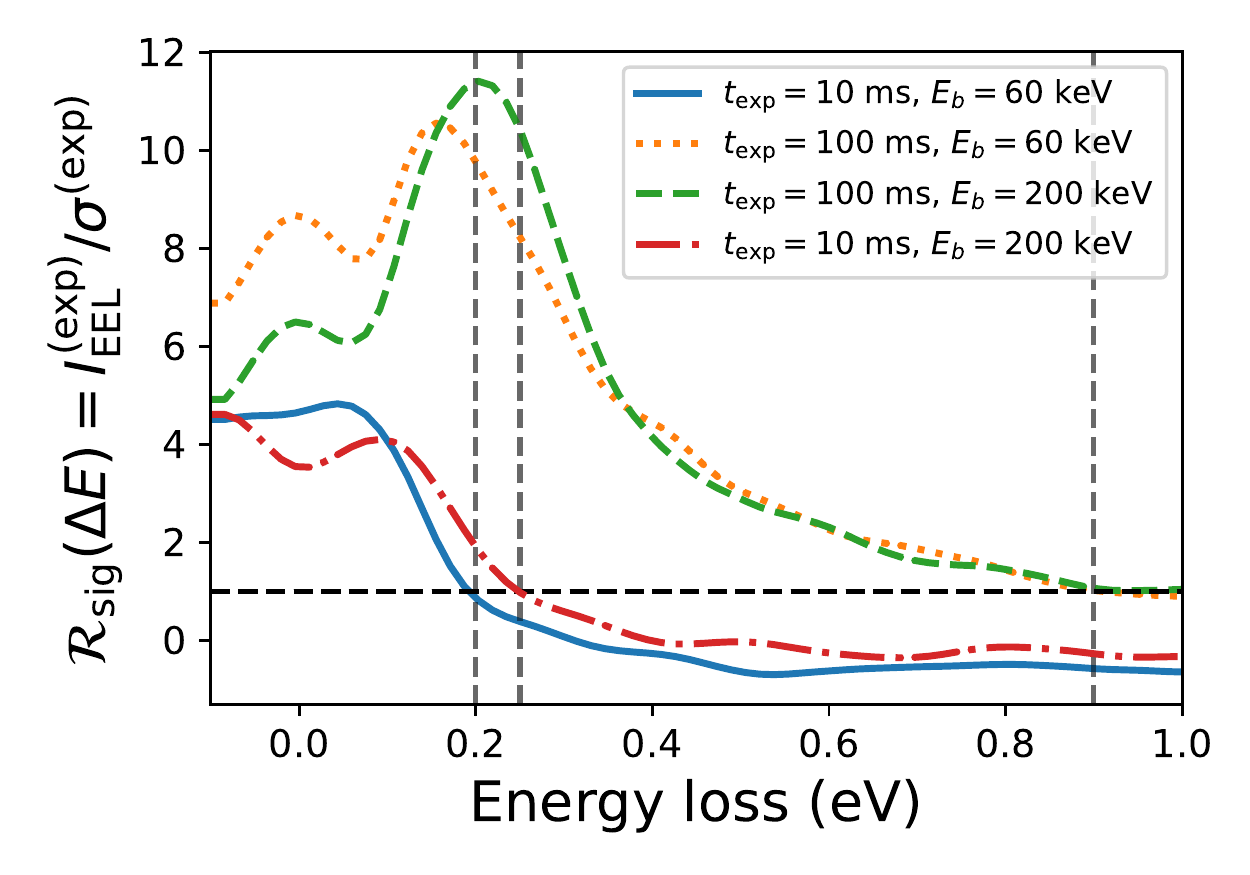}
  \caption{\small The ratio $\mathcal{R}_{\rm sig}(\Delta E)$
    between the central experimental value of the 
    EELS intensity distribution and its corresponding
    uncertainty, Eq.~(\ref{eq:pdlocation}).
    Results are shown for the four combinations of $t_{\rm exp}$
    and $E_{b}$ listed in Table~\ref{table:vacuumdata}.
    The vertical dashed lines mark the values of $\Delta E$ for which
    $\mathcal{R}_{\rm sig}\simeq 1$, which indicates when the
    data is dominated by statistical noise.
  }
  \label{fig:intensityratio}
\end{figure}
%%%%%%%%%%%%%%%%%%%%%%%%%%%%%%%%%%%%%%%%%%%%%%%%%%%%%%%%%%%%%%%%%%%%%%%%%%%%%%%%%%%%%%%%%%%%%%%%%%%%%%%%%%%%%%%%%

Fig.~\ref{fig:intensityratio} displays this ratio
for the four combinations of $t_{\rm exp}$
and $E_{b}$ listed in Table~\ref{table:vacuumdata}.
The vertical dashed lines indicate the values of $\Delta E$ for which
$\mathcal{R}_{\rm sig}$ becomes smaller than unity.
For larger $\Delta E$, the EELS spectra become
consistent with zero within uncertainties and can thus be discarded and replaced
by the pseudo-data constraints.
The total uncertainty of the pseudo-data points is then chosen to be
\be
\sigma_j^{(\rm pd)} = \frac{1}{10}I_{{\rm EEL}}^{\rm (exp)}\lp \Delta E = \Delta E_{\rm pd}^{\rm (min)}\rp \,, \quad 
j= 1,\ldots,N_{\rm pd} \, .
\ee
The factor of 1/10 is found to be suitable to ensure that the constraint
is enforced without distorting
the training to the experimental data.
We observe from Fig.~\ref{fig:intensityratio} that $\Delta E_{\rm pd}^{\rm (min)}$ depends
the operation conditions, with $\Delta E_{\rm pd}^{\rm (min)} \simeq 200$ meV for $t_{\rm exp}=10$ ms
and $\simeq  900$ meV for 100 ms, roughly independent on the value of the beam energy $E_b$.

The input experimental measurements listed in Table~\ref{table:vacuumdata} are used
to generate a sample of $N_{\rm rep}=500$ Monte Carlo replicas
and to train an individual neural network to each of these replicas.
The end result of the procedure is a set of model replicas,
\be
\label{eq:modelreplicas}
I_{\rm ZLP}^{\rm (mod)(k)}(\Delta E, E_{b},t_{\rm exp}) \, , \quad k=1,\ldots,N_{\rm rep} \, ,
\ee
which can be used to provide a prediction for the intensity of the ZLP
for arbitrary values of $\Delta E$,  $E_{b}$, and $t_{\rm exp}$.
Eq~(\ref{eq:modelreplicas})
provides the sought-for representation of the probability density in the space of ZLP models.
By means of this sample of replicas, one can evaluate
statistical estimators such as averages, variances, and correlations (as well
as higher moments) as follows:
\be
\label{eq:average}
\la I_{\rm ZLP}^{\rm (mod)}( \{z_1\}) \ra = \frac{1}{N_{\rm rep}}\sum_{k=1}^{N_{\rm rep}}
I_{\rm ZLP}^{\rm (mod)(k)}( \{z_1\}) \, ,
\ee
\be
\label{eq:standarddev}
\sigma_{I_{\rm ZLP}}^{\rm (mod)}( \{z_1\})  = \lp \frac{1}{N_{\rm rep}-1} \sum_{k=1}^{N_{\rm rep}}
\lp  I_{\rm ZLP}^{\rm (mod)(k)}  - \la I_{\rm ZLP}^{\rm (mod)}  \ra   \rp \rp^{1/2} \, ,
\ee
\be
\rho \lp \{z_1\},\{z_2\}\rp = \frac{ \la I_{\rm ZLP}^{\rm (mod)}( \{z_1\} ) I_{\rm ZLP}^{\rm (mod)}( \{z_2\} ) \ra
- \la I_{\rm ZLP}^{\rm (mod)}( \{z_1\} )\ra \la I_{\rm ZLP}^{\rm (mod)}( \{z_2\} ) \ra}{\sigma_{I_{\rm ZLP}}^{\rm (mod)}( \{z_1\} )\sigma_{I_{\rm ZLP}}^{\rm (mod)}( \{z_2\} )} \, ,
\ee
where as in the previous section $\{z_l\}$ denotes a possible set of input variables for the model,
here $\{z_l\}=\lp \Delta E_l, E_{b,l},t_{{\rm exp},l}\rp$.

\subsection{Fit quality}

We would like now to evaluate the overall fit quality of the neural network
model and demonstrate that it is flexible enough
to describe the available input datasets.
In Table~\ref{table:chi2summary} we indicate the values of the final $\chi^2$ per data point,
Eq.~(\ref{eq:chi2_final}), as well as the average values of the cost
function Eq.~(\ref{eq:chi2}) evaluated
over the training and validation subsets, for each of the four sets of spectra listed in
Table~\ref{table:vacuumdata} as well as for the total dataset.
We recall that for a satisfactory training one expects $\chi^2 \simeq 1$
and $\la E_{\rm tr}\ra \simeq \la E_{\rm val}\ra \simeq 2 $~\cite{Forte:2002fg}.
From the results of this table we find that, while our values
are  consistent with a reasonably good training,
somewhat lower values than expected are obtained,
for instance $\chi^2_{\rm tot}\simeq 0.8$ for the total dataset.
This suggests that correlations between the input data points might be partially missing, since neglecting
them often results into a moderate overestimate of the experimental uncertainties.

%%%%%%%%%%%%%%%%%%%%%%%%%%%%%%%%%%%%%%%%%%%%%%%%%%%%%%%%%%%%%%%%%%%%%%%%%%%%%%%%%%%%%%%%%%%%%
%%%%%%%%%%%%%%%%%%%%%%%%%%%%%%%%%%%%%%%%%%%%%%%%%%%%%%%%%%%%%%%%%%%%%%%%%%%%%%%%%%%%%%%%%%%%%
\begin{table}[t]
  \begin{center}
            \renewcommand{\arraystretch}{1.35}
  \begin{tabular}{@{}cccc}
\br
$\quad$Set$\quad$ & $\qquad \chi^2\qquad$  &  $\qquad\la E_{\rm tr}\ra\qquad$   &  $\qquad\la E_{\rm val}\ra\qquad$ \\
\mr
1        &           1.00        &      1.70            &  1.97  \\
2        &           0.73        &     1.41            &  1.77  \\
3        &           0.70        &    1.39            &  1.80  \\
4        &           0.60        &    1.20            &  1.76  \\
\mr
Total    &           0.77        &    1.47           &  1.85  \\
\br
  \end{tabular}
    \end{center}
  \caption{\small \small The values of the $\chi^2$ per data point,
    Eq.~(\ref{eq:chi2_final}), as well as the average values of the cost function Eq.~(\ref{eq:chi2})
    over the training $\la E_{\rm tr}\ra$ and validation $\la E_{\rm val}\ra$ subsets, for each of the four sets of spectra listed in
    Table~\ref{table:vacuumdata} as well as for the total dataset used in the present analysis.
  }
   \label{table:chi2summary}
\end{table}
%%%%%%%%%%%%%%%%%%%%%%%%%%%%%%%%%%%%%%%%%%%%%%%%%%%%%%%%%%%%%%%%%%%%%%%%%%%%%%%%%%%%%%%%%%%%%%%%%5
%%%%%%%%%%%%%%%%%%%%%%%%%%%%%%%%%%%%%%%%%%%%%%%%%%%%%%%%%%%%%%%%%%%%%%%%%%%%%%%%%%%%%%%%%%%%%

\begin{figure}[t]
    \centering
    \includegraphics[width=120mm]{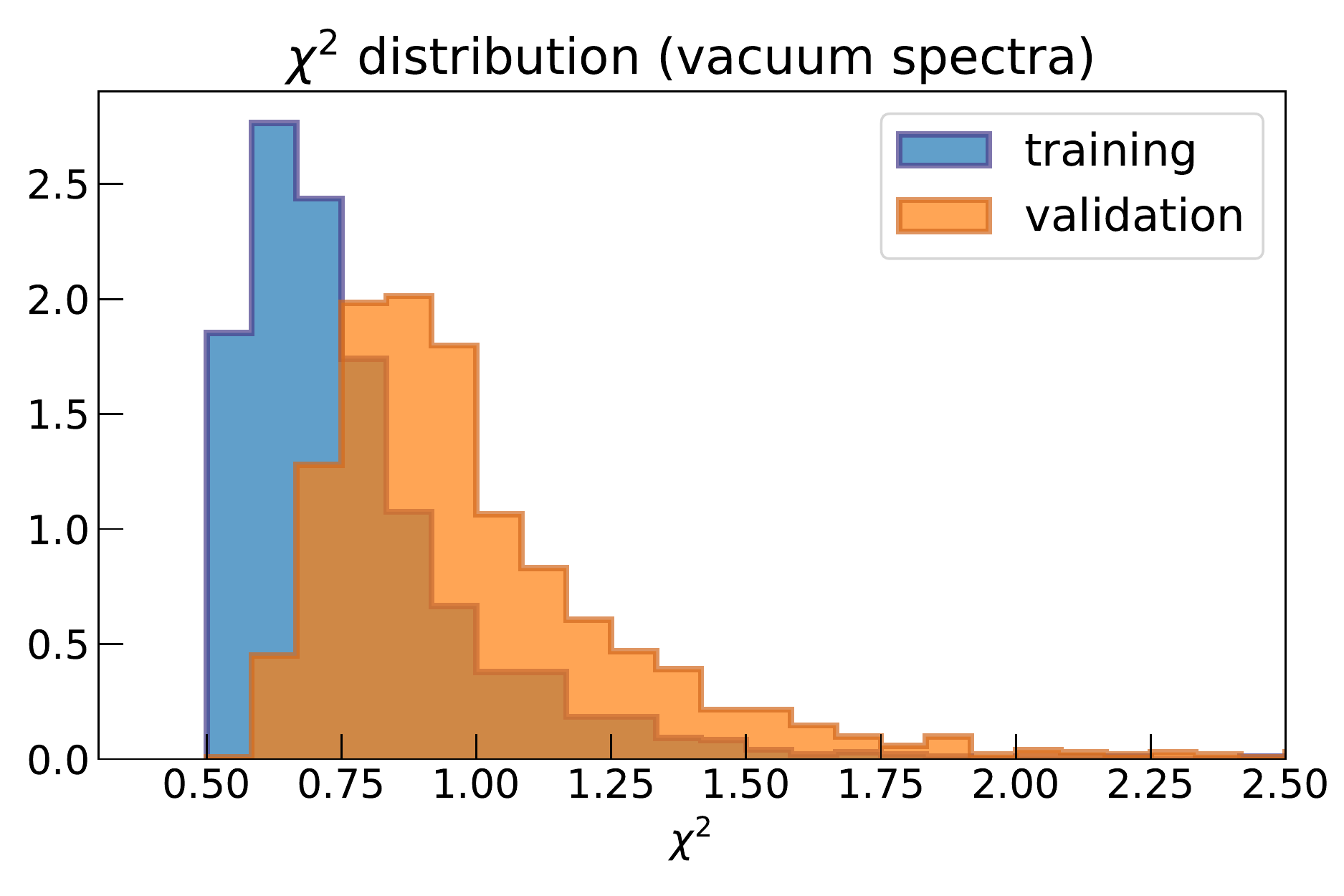}
    \caption{\small The distribution of the $\chi^2$ per data point evaluated
      separately for the training and validation sets over
      the $N_{\rm rep}=500$ replicas trained on the spectra
      listed in Table~\ref{table:vacuumdata}.}
    \label{fig:chi2_distributions}
\end{figure}

Then Fig.~\ref{fig:chi2_distributions} displays separately the $\chi^2$  distributions
evaluated for the training and validation sets
of the $N_{\rm rep}=500$ replicas of the sample trained on the spectra
listed in Table~\ref{table:vacuumdata}.
Note that the training/validation partition differs at random for each replica.
The $\chi^2_{\rm tr}$ distribution peaks at $\chi^2_{\rm tr}\simeq 0.7$,
indicating that a satisfactory model training
has been achieved, but also that the errors on the input data points might have
been slightly overestimated.
We emphasize that the stopping criterion for the neural net training adopted here never considers
the absolute values of the error function and determines proper learning entirely from
the global minima of $E_{\rm val}^{(k)}$.
From Fig.~\ref{fig:chi2_distributions} we also observe that  the validation
distribution peaks at
a slighter higher value, $\chi^2_{\rm val}\simeq 1$, and
is broader that its corresponding training counterpart.
These results confirm both that a satisfactory model training that prevents overlearning
has been achieved as well as an appropriate estimate of the statistical uncertainties
associated to the original EEL spectra.

%%%%%%%%%%%%%%%%%%%%%%%%%%%%%%%%%%%%%%%%%%%%%%%%%%%%%%%%%%%%%%%%%%%%%%%%%%%%%%%%%%%%%%%%%%%%%%%%%%%%%%%%%%%%
%%%%%%%%%%%%%%%%%%%%%%%%%%%%%%%%%%%%%%%%%%%%%%%%%%%%%%%%%%%%%%%%%%%%%%%%%%%%%%%%%%%%%%%%%%%%%%%%%%%%%%%%%%%%%%%%%%%

\subsection{Dependence on the electron energy loss}
\label{eq:depdeltae}

\begin{figure}[t]
    \centering
    \includegraphics[width=170mm]{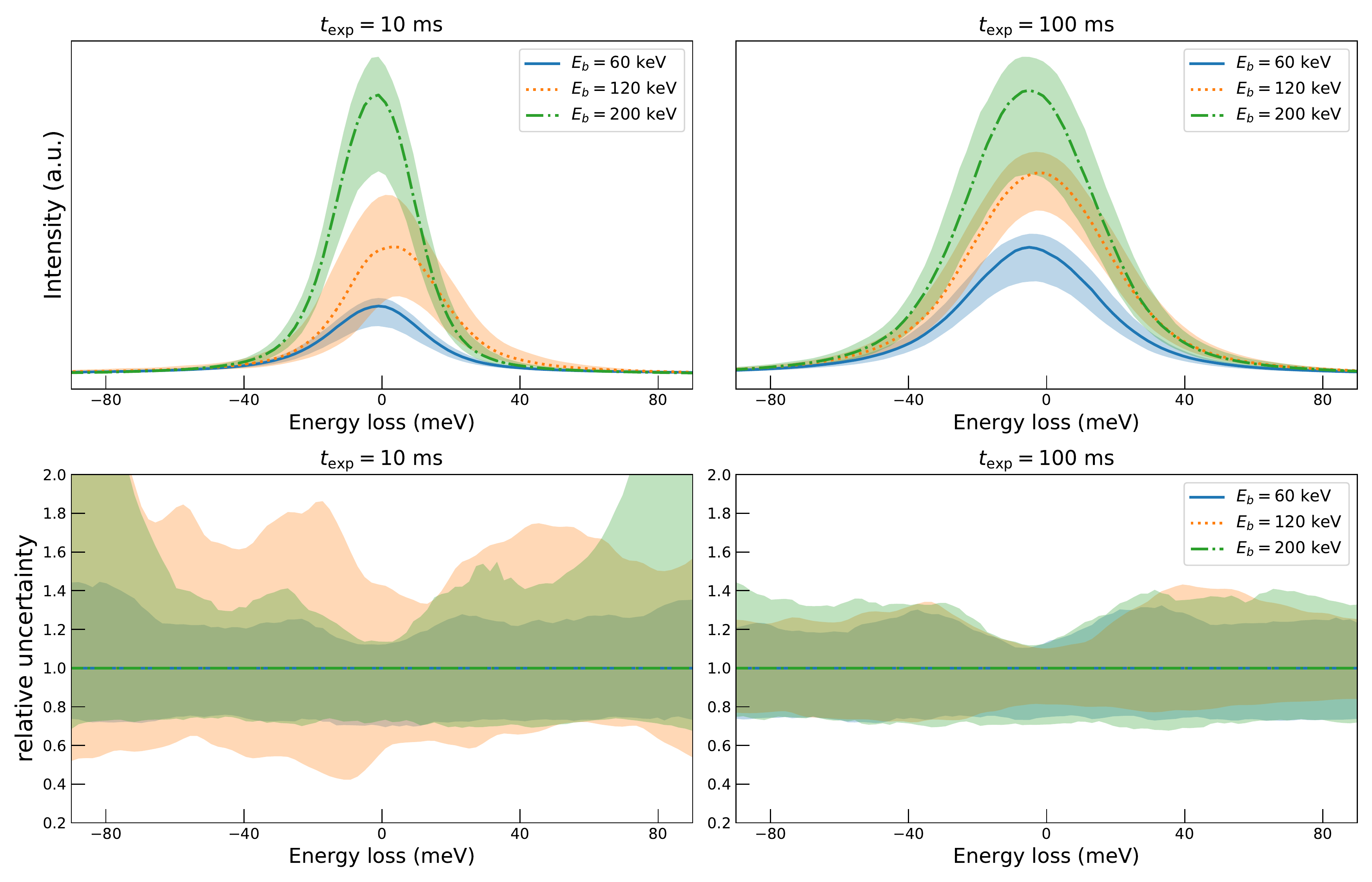}
    \caption{\small Top: the central value and 68\% confidence level uncertainty band
      for the ZLP model as a function
      of electron energy loss $\Delta E$
      evaluated using Eqns.~(\ref{eq:average}) and~(\ref{eq:standarddev}).
      We display results corresponding to 
      three different values of $E_b$  and for both
      $t_{\rm exp}=10$ ms (left)  and $t_{\rm exp}=100$ ms (right panel).
      Note that no training data with $E_b=120$ keV has been used and thus our prediction
      in that case arises purely from the model interpolation.
      Bottom: the corresponding relative uncertainty as a function of $\Delta E$
      for each of the three values of $E_b$.
      \label{fig:EELS_vacuum_DeltaE}}
\end{figure}

Having demonstrated that our neural network model provides a satisfactory description
of the input EEL spectra, we now present its  predictions for specific
choices of the input parameters.
First of all, we investigate the dependence of the results as a function of the
electron energy loss.
Fig.~\ref{fig:EELS_vacuum_DeltaE} displays the central value and 68\% confidence level uncertainty band
for the ZLP model as a function
of electron energy loss $\Delta E$
evaluated using Eqns.~(\ref{eq:average}) and~(\ref{eq:standarddev}).
We display results corresponding to 
three different values of $E_b$  and for both
$t_{\rm exp}=10$ ms and  $100$ ms.
We emphasize that no measurements with $E_b=120$ keV have been used in the training and thus our prediction
in that case arises purely from the model interpolation.
It is interesting to note how both the overall normalisation and the shape of
the predicted ZLP depend on the specific operating conditions.

%%%%%%%%%%%%%%%%%%%%%%%%%%%%%%%%%%%%%%%%%%%%%%%%%%%%%%%%%%%%%%%%%%%%%%%%%%%%%%%%%%%%%%%
%%%%%%%%%%%%%%%%%%%%%%%%%%%%%%%%%%%%%%%%%%%%%%%%%%%%%%%%%%%%%%%%%%%%%%%%%%%%%%%%%%%%%%%%%

In the bottom panels of Fig.~\ref{fig:EELS_vacuum_DeltaE} we show
the corresponding relative uncertainties as a function of $\Delta E$
for each of the three values of $E_b$.
Recall that in this work we allow for non-Gaussian distributions and thus the central
value is the median of the distribution and the error band in general will
be asymmetric.
In the case of the $t_{\rm exp}=10$ ms results, we see how the model prediction
at $E_b=120$ keV typically exhibits larger uncertainties than the predictions
for the two values of $E_b$ for which we have training data.
In the case of $t_{\rm exp}=100$ ms instead, the model predictions display very similar
uncertainties for the three values of $E_b$, which furthermore depend only mildly on $\Delta E$.
One finds there that the uncertainties associated to the ZLP model are $\simeq 20\%$
for $|\Delta E| \lsim 100~{\rm meV}$.

\begin{figure}[t]
    \centering
    \includegraphics[width=0.49\textwidth]{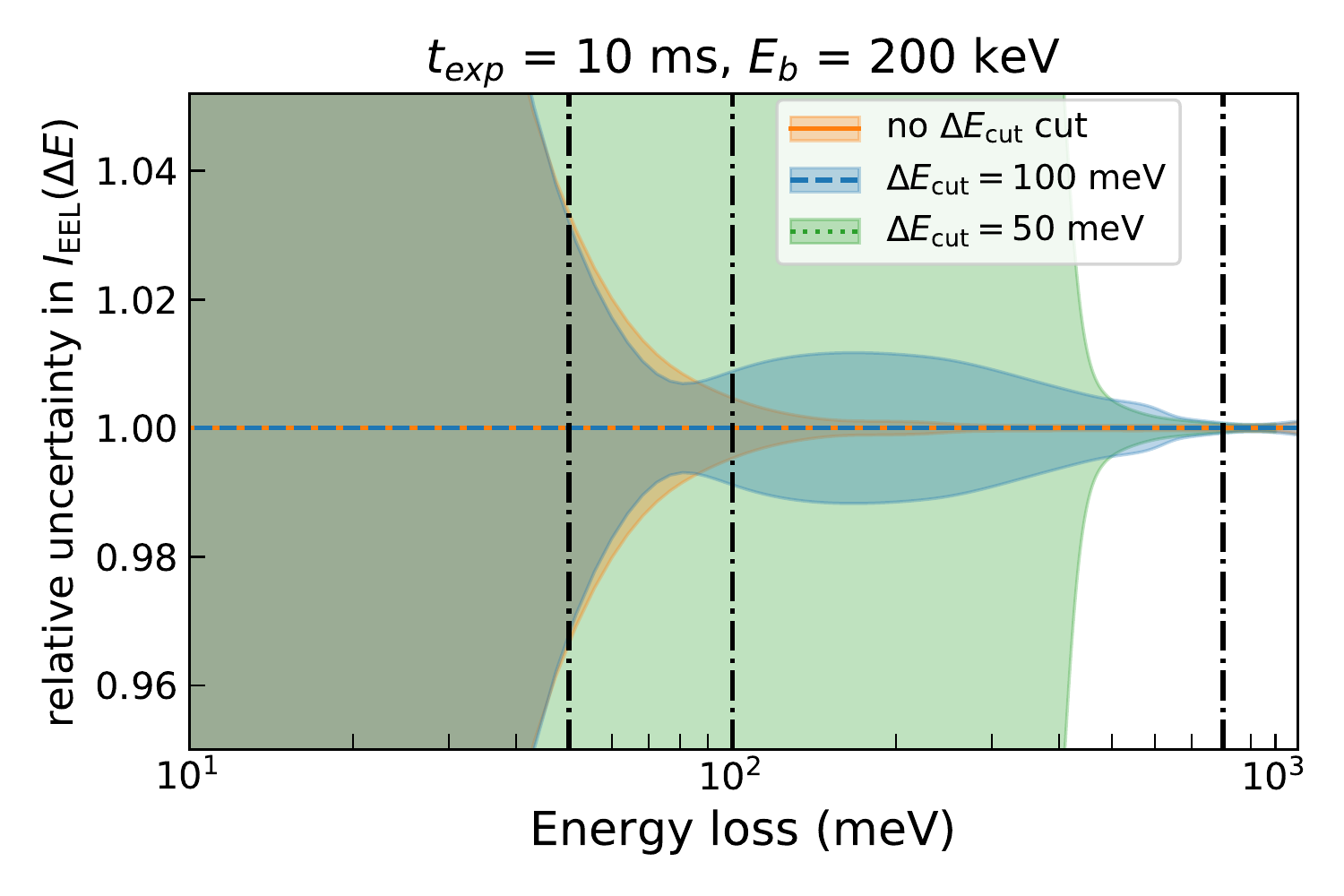}
    \includegraphics[width=0.49\textwidth]{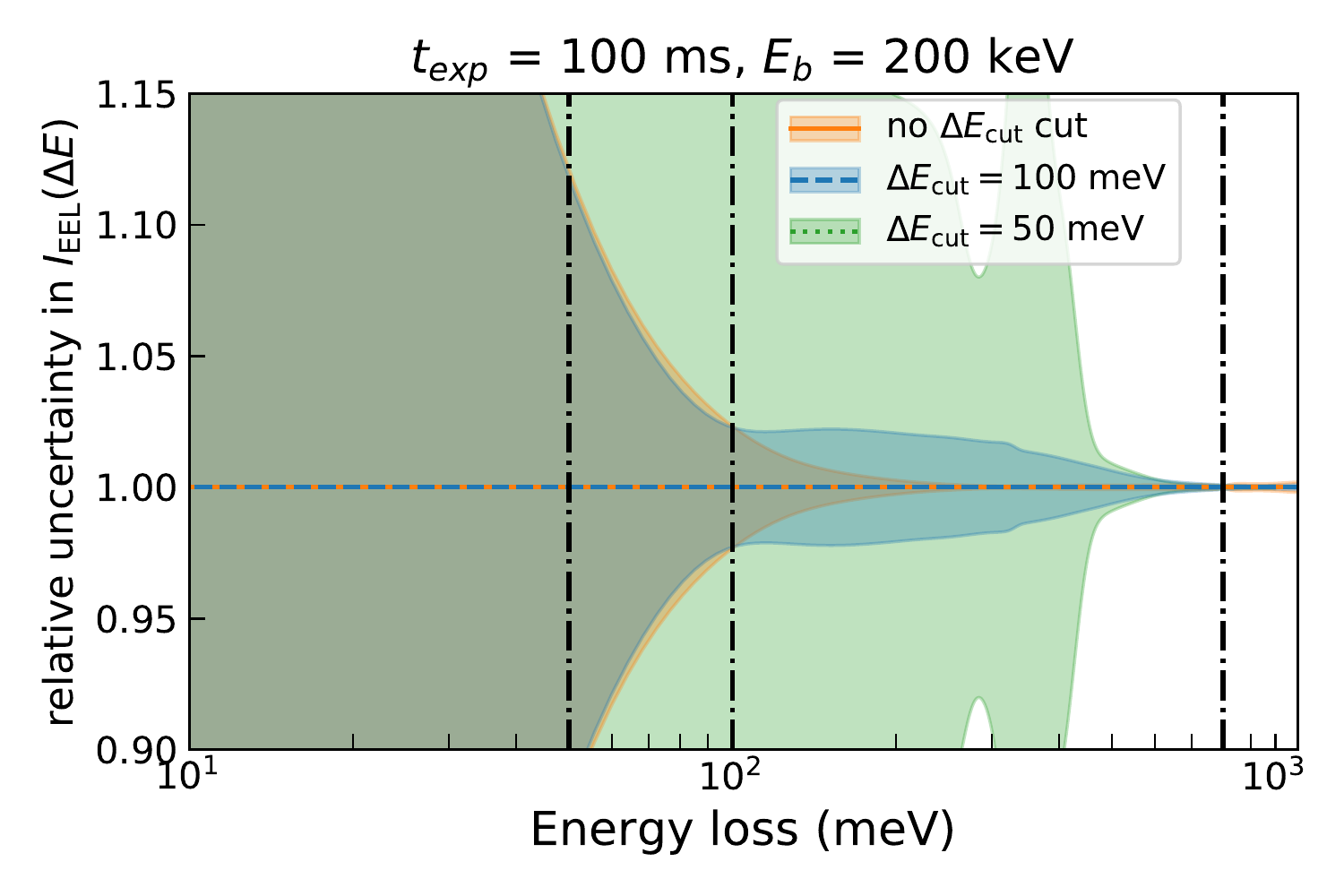}
    \caption{\small The relative uncertainty in the model predictions for $I_{\rm EEL}(\Delta E)$
      as a function of the energy loss for $E_b=200$ keV and $t_{\rm exp}=10$ ms (left)
      and 100 ms (right panel).
      We show results for three different cases: without any cut
      in the training dataset, and where the data points with $\Delta E \ge \Delta E_{\rm cut}$
      are removed from the training dataset for two different values
      of $\Delta E_{\rm cut}$.
      The same pseudo-data points that enforce $I_{\rm EEL}(\Delta E)\to 0$ are present
      in all three cases.
      \label{fig:EELS_vacuum_DeltaE_unc}}
\end{figure}

For the purpose of the second part of this work, it is important
to assess how the model results are modified once a subset of the data points
are excluded from the fit.
As illustrated in Fig.~\ref{fig:EELS_toy}, when training the model on sample spectra, the
region defined by
with $\Delta E_{\rm I} \le \Delta E \le \Delta E_{\rm II}$ will be removed from the training dataset to avoid the
contamination from the inelastic contributions.
To emulate the effects of such cut, 
Fig.~\ref{fig:EELS_vacuum_DeltaE_unc} displays
the relative uncertainty in the model predictions for $I_{\rm ZLP}(\Delta E)$
as a function of the energy loss for $E_b=200$ keV and $t_{\rm exp}=10$ ms 
and 100 ms.
We show results for three different cases: first of all, one without any cut
in the training dataset, and then for two cases where data points with $\Delta E \ge \Delta E_{\rm cut}$
are removed from the training dataset.
We consider two values of $\Delta E_{\rm cut}$, namely 50 meV and 100 meV, indicated
with vertical dash-dotted lines.
In both cases, data points are removed up until $\Delta E =$ 800 meV. The pseudo-data points 
that enforce the $I_{\rm EEL}(\Delta E)\to 0$ condition are present
in all three cases in the region 800 meV $\le \Delta E \le 1~{\rm eV}$. 

%%%%%%%%%%%%%%%%%%%%%%%%%%%%%%%%%%%%%%%%%%%%%%%%%%%%%%%%%%%%%%%%%%%%%%%%%%%%%%
%%%%%%%%%%%%%%%%%%%%%%%%%%%%%%%%%%%%%%%%%%%%%%%%%%%%%%%%%%%%%%%%%%%%%%%%%%%%%%%%

From this comparison one can observe how the model predictions become markedly more uncertain
once a subset of the training data is cut away, as expected due to the effect of the information
loss.
While for the cut $\Delta E_{\rm cut}=100$ meV the increase in model uncertainty is only moderate
as compared with the baseline fit where no cut is performed (since for this value of $\Delta E$
uncertainties are small to begin with), rather more dramatic effects are observed
for a value of the cut $\Delta E_{\rm cut}=50$ meV.
This comparison highlights how ideally we would like to keep as many data points
in the training set for the ZLP model, provided of course one can verify that the
possible contributions to the spectra related to inelastic scatterings from the
sample can be neglected.

\subsection{Dependence on beam energy and exposure time }
\label{eq:depebeam}

\begin{figure}[t]
    \centering
    \includegraphics[width=0.49\textwidth]{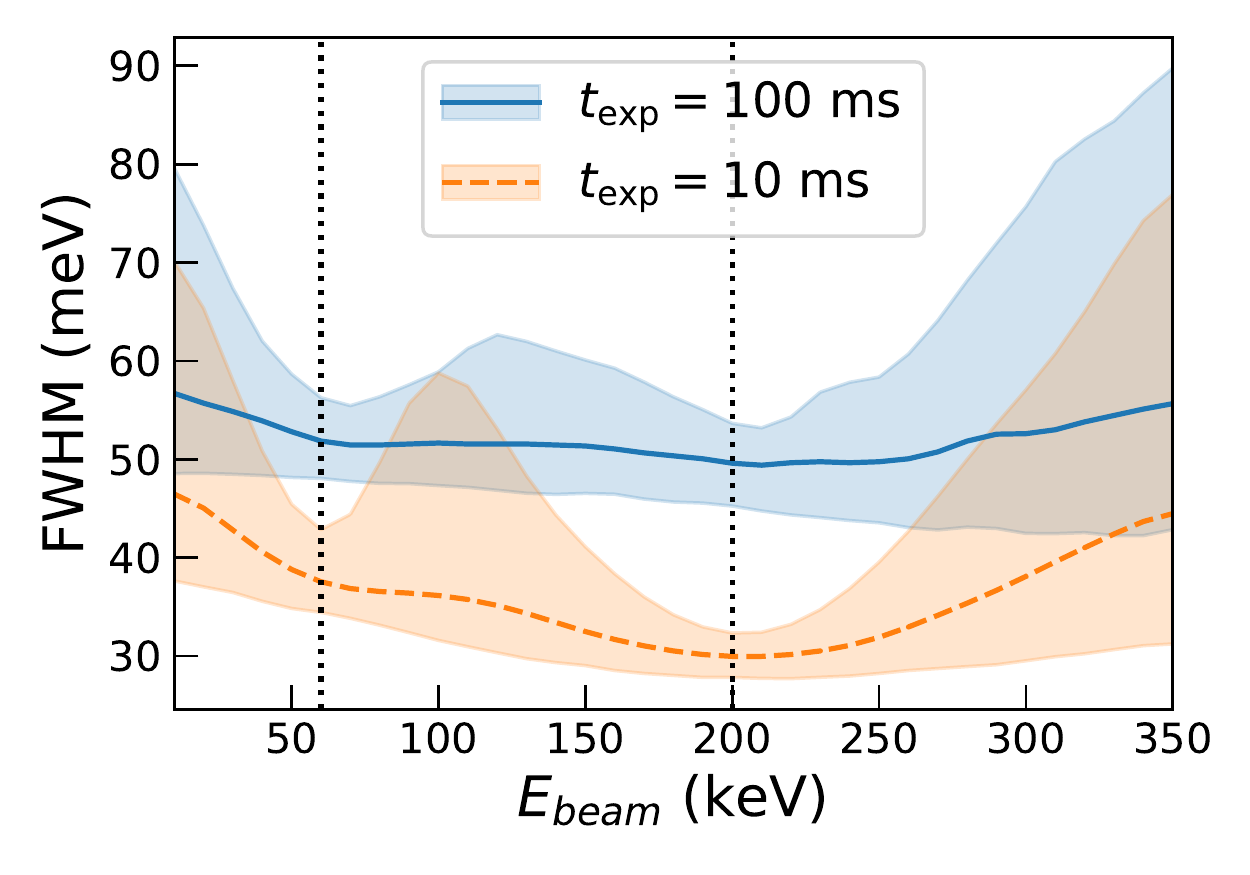}
    \includegraphics[width=0.49\textwidth]{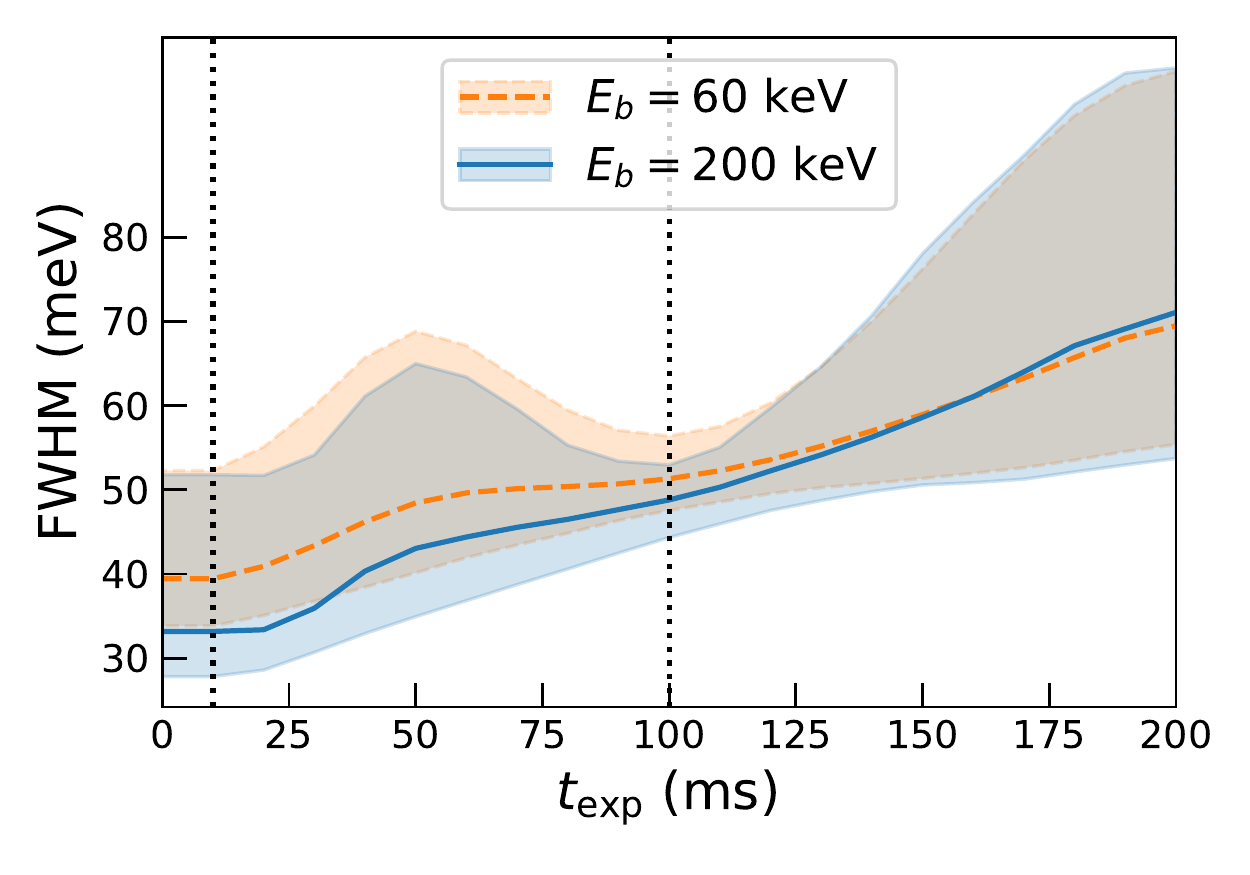}
    \caption{\small The model predictions for the FWHM of the zero-loss peak
      with its corresponding uncertainty as a function of the beam energy $E_b$
      for two values of the exposure time (left panel)
      and as a function of $t_{\rm exp}$ for two values of $E_b$ (right panel).
      The vertical dashed lines indicate the values of the
      corresponding microscope operation parameter for which we have training data.
    }
    \label{fig:extrapolbeam}
\end{figure}

As indicated in Table~\ref{table:vacuumdata}, the training dataset contains
spectra taken at two values of the electron beam energy, $E_b=60$ keV and 200 keV.
The left panel of Fig.~\ref{fig:extrapolbeam} displays the predictions for the FWHM of the zero-loss peak
(and its corresponding uncertainty) as a function of the beam energy $E_b$
for two values of the exposure time, $t_{\rm exp}=10$ ms and 100 ms.
The vertical dashed lines indicate the two values of $E_b$ for which spectra
are part of the training dataset.
This comparison illustrates how the model uncertainty differs between the data region
(near $E_b=60$ keV and 200 keV), the interpolation region (for $E_b$ between 60 and 200 keV),
and the extrapolation regions (for $E_b$ below 60 keV and above 200 keV).
In the case of $t_{\rm exp}=100$ ms for example, we observe that the model interpolates reasonably well
between the measured values of $E_b$ and that uncertainties increase
markedly in the extrapolation region above $E_b=200$ keV.
      
%%%%%%%%%%%%%%%%%%%%%%%%%%%%%%%%%%%%%%%%%%%%%%%%%%%%%%%%%%%%%%%%%%%%%%%%%%%%%%%%%%%%%%%
%%%%%%%%%%%%%%%%%%%%%%%%%%%%%%%%%%%%%%%%%%%%%%%%%%%%%%%%%%%%%%%%%%%%%%%%%%%%%%%%%%%%%%%%%%%%

From this comparison one can also observe how as expected the uncertainty in the  prediction for
the FWHM of the ZLP is the smallest close to the values of $E_b$ for which one has training data.
The uncertainties increase but only in a moderate way in the interpolation region, indicating that
the model can be applied to reliably predict the features of the ZLP for other values of the electron
energy beam (assuming that all other operation conditions of the microscope are unchanged).
The errors then increase rapidly in the extrapolation region, which is a characteristic
(and desirable) feature of  neural network models.
Indeed, as soon as the model departs from the data region there exists a very large
number of different functional form models for $I_{\rm ZLP}(\Delta E)$ that can describe equally well
the training dataset, and hence a blow up of the extrapolation uncertainties is generically expected.

In the same way as for the case of the electron beam energy $E_b$, while our ZLP model
was trained on data with only exposure times of $t_{\rm exp}=10$ and $100$ ms,
it can be used to reliably inter- and extrapolate to other values of $t_{\rm exp}$.
The right panel of Fig.~\ref{fig:extrapolbeam} displays the same
comparison as in the left one now as a function of $t_{\rm exp}$ for
$E_b=60$ keV and $E_b=200$ keV.
We observe that the FWHM increases approximately in a linear manner with the exposure time, indicating
that lower values of $t_{\rm exp}$ allow for an improved spectral resolution, and that the model
predictions are approximately independent of $E_b$.
Similarly to the predictions for varying beam energies, also for the exposure time the uncertainties grow bigger as the value of this parameter deviates more from the training inputs,
specially for large values of $t_{\rm exp}$.

All in all, we conclude that the predictions of the ML model trained on vacuum spectra
behave as they ought to: the smallest uncertainties correspond to the parameter values
that are included in the training dataset, while the largest uncertainties arise
in the extrapolation regions when probing regions of the parameter space far from those
present in the training set.

\section{Mapping low-loss EELS in polytypic WS$_2$}
\label{sec:results_sample}

Following the discussion of the vacuum ZLP analysis, we now
present the application of our machine learning strategy to parametrise the ZLP
arising in spectra recorded on specimens, specifically for
EELS measurements acquired in different regions
of the WS$_2$ nanoflowers presented in Sect.~\ref{sec:tmd}.
The resulting ZLP parametrisation will be applied to isolate the inelastic
contribution in each spectrum.
We will use these subtracted spectra first to determine the bandgap type and energy 
value from the behaviour of the onset region and second to identify excitonic
transitions at very low energy losses.

In this section we begin by presenting the training dataset, composed by two groups of EEL spectra recorded
in thick and thin regions of the WS$_2$  nanoflowers respectively.
Then we discuss the subtraction procedure, the choice of hyper-parameters, and the error propagation
to the physical predictions.
The resulting subtracted spectra provide the information
required to extract the value and type of the bandgap
and to characterise excitonic transitions for different regions of these polytypic WS$_2$ nanostructures.

\subsection{Training dataset}

\begin{figure}[t]
\begin{centering}
  \includegraphics[width=0.92\linewidth]{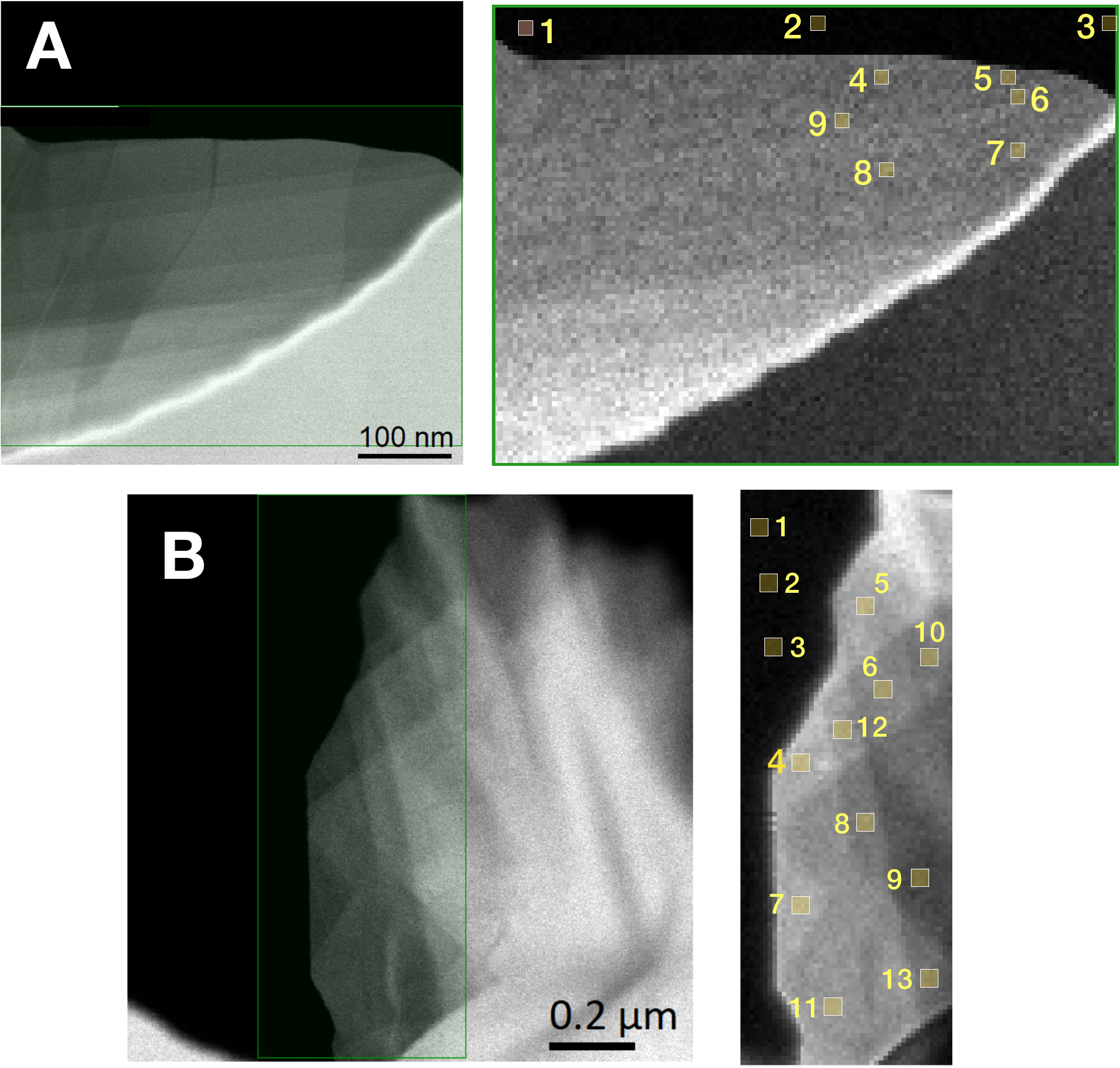}
  \caption{Low-magnification TEM images (left) and the corresponding
    spectral images (right panels) of two different regions of
    the WS$_2$ nanoflowers, denoted as sample A (upper) and sample B (lower panels) respectively.
    The spectral images have been recorded in the regions marked by a green square
    in the associated TEM images, and contain an individual EEL spectrum in each pixel.
    We indicate the locations where representative
    EEL spectra have been selected. 
    In the left panel of sample B, the difference in contrast is correlated to the material
    thickness, with higher contrast indicating thinner regions of the nanostructure.
    The morphological differences between the two samples are discussed in the text.
  }
\label{fig:ws2positions}
\end{centering}
\end{figure}

Low-magnification TEM images  and the corresponding
spectral images of two representative regions of
the WS$_2$ nanoflowers, denoted as sample A and B  respectively, are displayed in Fig.~\ref{fig:ws2positions}.
These spectral images have been recorded in the regions marked by a green square
in the associated TEM images, and contain an individual EEL spectrum in each pixel.
We indicate the specific locations where
EEL spectra have been recorded, including the in-vacuum measurements acquired
for calibration purposes.
Note that in sample B  the differences in contrast are related to the material
thickness, with higher contrast corresponding to thinner regions.

These two samples are characterised by rather different structural morphologies.
While sample A is composed
by a relatively thick region of WS$_2$, sample B corresponds to a region where thin petals
overlap between them.
In other words, sample A is composed by bulk WS$_2$ while in sample B some specific regions
could be rather thinner, down to the few monolayers level.
This thickness information has been be determined
by means of the {\tt Digital~Micrograph}  software.

%%%%%%%%%%%%%%%%%%%%%%%%%%%%%%%%%%%%%%%%%%%%%%%%%%%%%%%%%%%%%%%%%%%%%%%
%%%%%%%%%%%%%%%%%%%%%%%%%%%%%%%%%%%%%%%%%%%%%%%%%%%%%%%%%%%%%%%%%%%%%%%%%%

One of the main goals of this study is demonstrating that our ZLP-subtraction method exhibits
a satisfactory performance for spectra taken with different microscopes and operation conditions.
With this motivation, the EELS measurements acquired on specimens A and B have
been obtained varying both the microscopes and their settings.
Specifically, the TEM and EELS measurements acquired in specimen A  are based on a JEOL 2100F
microscope with a cold field-emission
gun and equipped with an aberration corrector,
operated at 60 kV and where a Gatan GIF Quantum was used for
the EELS analysis.
The corresponding measurements on specimen B  were recorded instead
using a JEM ARM200F monochromated microscope operated at 60 kV and equipped with a GIF quantum ERS.
See Methods for more details.

In Table~\ref{table:sampledata} we collect the most relevant properties of the spectra collected
in the locations indicated in Fig.~\ref{fig:ws2positions} using the same format as
in Table~\ref{table:vacuumdata}.
As we just mentioned, the spectra from samples A and B
have been acquired with different microscopes and thus features of the ZLP
such as the FWHM are expected to be different.
From this table one can observe how the ZLP for the spectra acquired on sample A exhibit
a FWHM about five times larger as compared to those of sample B.
This difference in energy resolution can be understood from the fact that the EELS spectra from sample B, unlike those
from sample A, were recorded with a TEM equipped with monochromator.

%%%%%%%%%%%%%%%%%%%%%%%%%%%%%%%%%%%%%%%%%%%%%%%%%%%%%%%%%%%%%%%%%%%%%%%%%%%%%%%%%%%%%%%%%%%%%
%%%%%%%%%%%%%%%%%%%%%%%%%%%%%%%%%%%%%%%%%%%%%%%%%%%%%%%%%%%%%%%%%%%%%%%%%%%%%%%%%%%%%%%%%%%%%
\begin{table}[t]
  \begin{center}
            \renewcommand{\arraystretch}{1.50}
  \begin{tabular}{@{}ccccccccc}
\br
Set & $t_{\rm exp}$ {(}ms{)} & $E_{\rm b}$ {(}keV{)} & $N_{\rm sp}$ & $N_{\rm dat}$ & $\Delta E_{\rm min}$~(eV)  & $\Delta E_{\rm max}$~(eV)  & FWHM~(meV)  \\ 
\mr
A        &       1       &        60         &   6      &    1918    &     -4.1       & 45.5 & $ 470\pm 10 $  \\
B        &       190       &        60       &   10     &    2000    &     -0.9        & 9.1   & $ 87 \pm 5$ \\
\br
  \end{tabular}
    \end{center}
  \caption{\small Same as Table~\ref{table:vacuumdata} for the EEL spectra taken on specimens A and B.
    The location on the WS$_2$ nanoflowers where each spectra has been recorded
    is indicated in Fig.~\ref{fig:ws2positions}.
  }
   \label{table:sampledata}
\end{table}
%%%%%%%%%%%%%%%%%%%%%%%%%%%%%%%%%%%%%%%%%%%%%%%%%%%%%%%%%%%%%%%%%%%%%%%%%%%%%%%%%%%%%%%%%%%%%%%%%5
%%%%%%%%%%%%%%%%%%%%%%%%%%%%%%%%%%%%%%%%%%%%%%%%%%%%%%%%%%%%%%%%%%%%%%%%%%%%%%%%%%%%%%%%%%%%%

In the following we will present results for representative spectra
corresponding to specific choices of the locations indicated in Fig.~\ref{fig:ws2positions}.
The full set of recorded spectra is available  within {\tt EELSfitter},
the code used to produce the results of this analysis, and
whose installation
and usage instructions are summarised in Appendix~\ref{sec:installation}.

\subsection{Subtraction procedure}

In Table~\ref{table:sampledata_summary} we collect
the mean value and uncertainty of the first local minimum, $\Delta E|_{\rm min}$.
averaged over the spectra corresponding to samples A and B from
Fig.~\ref{fig:ws2positions}.
The location of the first minimum is relatively stable
among all the spectra belonging to a given set.
This indicates that the onset of the inelastic contributions $I_{\rm inel}$ does
not change significantly as we move between different regions of the sample.
We also indicate there
the corresponding values of the hyper-parameters
$\Delta E_{\rm I}$ and $\Delta E_{\rm II}$ defined in Fig.~\ref{fig:EELS_toy}.
Recall that only
the data points with $\Delta E \le \Delta E_{\rm I}$ are used for the training
of the neural network model.
The model training is performed for a range of $\Delta E_{\rm I}$ values,
subject to the condition that $\Delta E_{\rm I} \le \Delta E_{\rm min}$, to validate
the stability of the results.
The optimal value of $\Delta E_{\rm I}$ is determined by the condition
that Eq.~(\ref{eq:rder}) satisfies $\mathcal{R}^{(j)}_{\rm der}(\Delta E)\simeq 0.9$, indicating
that the shape of the intensity profile for the sample spectra differs by more than 10\%
as compared to their vacuum counterparts.

%%%%%%%%%%%%%%%%%%%%%%%%%%%%%%%%%%%%%%%%%%%%%%%%%%%%%%%%%%%%%%%%%%%%%%%%%%%%%%%%%%%%%%%%%%%%%
%%%%%%%%%%%%%%%%%%%%%%%%%%%%%%%%%%%%%%%%%%%%%%%%%%%%%%%%%%%%%%%%%%%%%%%%%%%%%%%%%%%%%%%%%%%%%
\begin{table}[t]
  \begin{center}
            \renewcommand{\arraystretch}{1.50}
  \begin{tabular}{@{}ccccccccc}
\br
Set & $\Delta E|_{\rm min}$~(eV)  &  $\Delta E_{\rm I}$~(eV)  &  $\Delta E_{\rm II}$~(eV)   \\
\mr
A        &    $2.70\pm0.06$               &          1.8        &      12         \\
B        &    $1.80\pm0.04$               &          1.4        &      6        \\
\br
  \end{tabular}
    \end{center}
  \caption{\small The mean value and uncertainty of the first local minima, $\Delta E|_{\rm min}$,
    averaged over the spectra corresponding to samples A and B from
    Fig.~\ref{fig:ws2positions}.
    We also indicate
     the corresponding values of the hyper-parameters
     $\Delta E_{\rm I}$ and $\Delta E_{\rm II}$ defined in Fig.~\ref{fig:EELS_toy} used for the training
     of the neural network model.
  }
   \label{table:sampledata_summary}
\end{table}
%%%%%%%%%%%%%%%%%%%%%%%%%%%%%%%%%%%%%%%%%%%%%%%%%%%%%%%%%%%%%%%%%%%%%%%%%%%%%%%%%%%%%%%%%%%%%%%%%5
%%%%%%%%%%%%%%%%%%%%%%%%%%%%%%%%%%%%%%%%%%%%%%%%%%%%%%%%%%%%%%%%%%%%%%%%%%%%%%%%%%%%%%%%%%%%%

In the region $\Delta E \ge \Delta E_{\rm II}$, the training set includes only the pseudo-data
that implements the $I_{\rm ZLP}(\Delta E)\to 0$ constraint.
The values for $\Delta E_{\rm II}$ were determined from the spectra recorded in vacuum
following the same procedure as explained 
in Sect.~\ref{sec:results_vacuum}, based on requiring $\mathcal{R}_{\rm sig}(\Delta E_{\rm II})\lsim 1$.
We note that the values of $\Delta E_{\rm II}$ found now are significantly higher than
the ones obtained in Fig.~\ref{fig:intensityratio} for the vacuum case.
This difference could be ascribed to the fact that 
the vacuum spectra from samples A and B were recorded in proximity to the sample so that the influence of the specimen is still partially felt.

The end result of the  neural network training described in Sect.~\ref{sec:training} is
 a set of $N_{\rm rep}=500$ replicas
 parametrising the zero-loss peak,
\be
 I_{\rm ZLP}^{({\rm mod})(k)}(\Delta E) \, ,\quad  k=1,\ldots,N_{\rm rep} \, .
\ee
 Taking into account that we have $N_{\rm sp}$ individual spectra in each sample,  the ZLP
 subtraction is performed individually
 for each Monte Carlo replica,
 \be
 \label{eq:subtractedModelPrediction}
 I_{\rm inel}^{({\rm exp})(j,k)}(\Delta E) \equiv I_{\rm EEL}^{({\rm exp})(j)}(\Delta E) - I_{\rm ZLP}^{({\rm mod})(k)}(\Delta E)\, ,
 \quad \forall~N_{\rm rep} \, ,\quad j=1,\ldots,N_{\rm sp} \, ,
 \ee
 from which statistical estimators can be evaluated.
 For instance, the mean value for our model prediction for the $j$-th spectrum
 can be evaluated by averaging over the replicas,
 \be
 \la  I_{\rm inel}^{({\rm exp})(j)}\ra (\Delta E)
 = \frac{1}{N_{\rm rep}} \sum_{k=1}^{N_{\rm rep}}  I_{\rm inel}^{({\rm exp})(j,k)}(\Delta E) \, ,
 \quad j=1,\ldots,N_{\rm sp} \, ,
 \ee
 and likewise for the corresponding uncertainties and correlation coefficients.
 For large values of $\Delta E$,
 the model prediction reduces to the original spectra, since in that region
 the ZLP contribution vanishes,
 \be
 I_{\rm inel}^{({\rm exp})(j,k)}(\Delta E \gg \Delta E_{\rm I}) \to  I_{\rm EEL}^{{\rm (exp)}(j)}(\Delta E) \, ,\quad
 \forall~j,k \, .
 \ee
 
 For very small values of the energy loss, the contribution to the total
 spectra from inelastic scatterings is negligible
 and thus the subtracted model prediction Eq.~(\ref{eq:subtractedModelPrediction}) should
 vanish.
 However, this will not be the case in general since the neural network model is trained on
 the $N_{\rm sp}$ ensemble of spectra, rather that just on individual ones, and thus the expected
 $\Delta E \to 0$ behaviour will only be achieved within uncertainties rather than at the level of
 central values.
 To achieve the desired $\Delta E \to 0$ limit, we apply a matching procedure
 as follows.
 We introduce another hyper-parameter, $\Delta E_0 < \Delta E_{\rm I}$, such that
 one has for the $k$-th ZLP replica associated to the $j$-th spectrum the following
 behaviour:
 \bea
 \nonumber
 I_{\rm ZLP}^{({\rm mod})(j,k)}(\Delta E) &=& I_{\rm EEL}^{({\rm exp})(j)}(\Delta E) \, ,\quad \Delta E < \Delta E_0  \, ,\\
 I_{\rm ZLP}^{({\rm mod})(j,k)}(\Delta E) &=& I_{\rm EEL}^{{\rm (exp)}(j)} + \lp \xi_1^{(n_l)(k)}(\Delta E) -
 I_{\rm EEL}^{{\rm (exp)}(j)}(\Delta E)\rp  \times \mathcal{F} \, , \nonumber \quad 
 \Delta E_0 < \Delta E \le \Delta E_{\rm I} \, ,\\
 &&\mathcal{F}(\Delta E) = \exp\lp -\frac{\lp \Delta E - \Delta E_{\rm I} \rp^2 }{\lp \Delta E_0 - \Delta E_{\rm I} \rp^2 \delta^2} \rp  \, , \label{eq:matching} \\
 I_{\rm ZLP}^{({\rm mod})(j,k)}(\Delta E) &=& \xi_1^{(n_l)(k)}(\Delta E) \, , \quad \Delta E > \Delta E_{\rm I} \nonumber \, ,
 \eea
 where $\xi_1^{(n_l)(k)}$ indicates the output of the $k$-th neural network that parametrises
 the ZLP and $\delta$ is a dimensionless tunable parameter.
 In Eq.~(\ref{eq:matching}), $\mathcal{F}(\Delta E)$ represents a matching factor
 that ensures that the ZLP model prediction smoothly interpolates
 between $\Delta E=\Delta E_0$ (where $\mathcal{F}\ll 1$ and the original spectrum should
 be reproduced) and $\Delta E=\Delta E_{\rm I}$
 (where $\mathcal{F}=1$ leaving the neural network output unaffected).
 Here we adopt $\Delta E_0 = \Delta E_{\rm I} -0.5\,{\rm eV}$,  having verified
 that results are fairly independent of this choice.
 Taking into account the matching procedure, we can slightly modify Eq.~(\ref{eq:subtractedModelPrediction})
 to 
 \be
 \label{eq:subtractedModelPrediction2}
 I_{\rm inel}^{({\rm mod})(j,k)}(\Delta E) \equiv I_{\rm EELS}^{({\rm exp})(j)}(\Delta E) - I_{\rm ZLP}^{({\rm mod})(j,k)}(\Delta E)\, ,
 \quad \forall~N_{\rm rep} \, ,\quad j=1,\ldots,N_{\rm sp} \, .
 \ee
 The ensemble of ZLP-subtracted spectra obtained this way, $\{I_{\rm inel}^{({\rm mod})(j,k)} \} $,
 can then be used to reliably extract physical information from the low-loss region of the spectrum.
 
 \subsection{Bandgap analysis of polytypic 2H/3R WS$_2$}

 One particularly important application of the ZLP-subtracted spectra is to
 estimate the specimen bandgap in the region where
 they were acquired.
 Different approaches  have been put forward to evaluate $E_{\rm BG}$ from 
subtracted EEL spectra, \textit{e.g.} by means of the inflection point of the rising intensity or
a linear fit to the maximum positive slope~\cite{Schamm:2003}.
Here we will adopt the approach of~\cite{Rafferty:2000} where the behaviour
of $I_{\rm inel}(\Delta E)$ in the onset region is modeled as
\begin{equation}
  \label{eq:I1}
    I_{\rm inel}(\Delta E) \simeq  A \lp \Delta E-E_{\rm BG} \rp^{b} \, , \quad \Delta E \ge E_{\rm BG} \, ,
\end{equation}
and vanishes for $\Delta E < E_{\rm BG}$, where both the bandgap value
$E_{\rm BG}$ as well as the parameters $A$ and $b$ are extracted from the fit.
The exponent $b$ is expected to be $b\simeq 1/2~(\simeq 3/2)$ for a semiconductor material characterised
by a direct~(indirect) bandgap.
 For each of the $N_{\rm sp}$ spectra and the $N_{\rm rep}$ replicas
 we fit to Eq.~(\ref{eq:subtractedModelPrediction2}) the model Eq.~(\ref{eq:I1})
 within a range taken to be
 $\lc \Delta E_{\rm I} - 0.5~{\rm eV}, \Delta E_{\rm I} + 0.7~{\rm eV}\rc$.
 One ends up with $N_{\rm rep}$ values for
 the bandgap energy and fit exponent for each spectra,
 \be
 \Big \{ E_{\rm BG}^{(j,k)}, b^{(j,k)} \Big\} \, , \quad k=1,\ldots,N_{\rm rep} \, ,
 \quad j=1,\ldots,N_{\rm sp} \, ,
 \ee
 from which again one can readily evaluate their statistical estimators.
 In the following, we will display the median and the 68\% confidence level intervals
 for these parameters to account for the fact that their distribution will be in general non-Gaussian.

\begin{figure}[t]
\begin{centering}
  \includegraphics[width=0.99\linewidth]{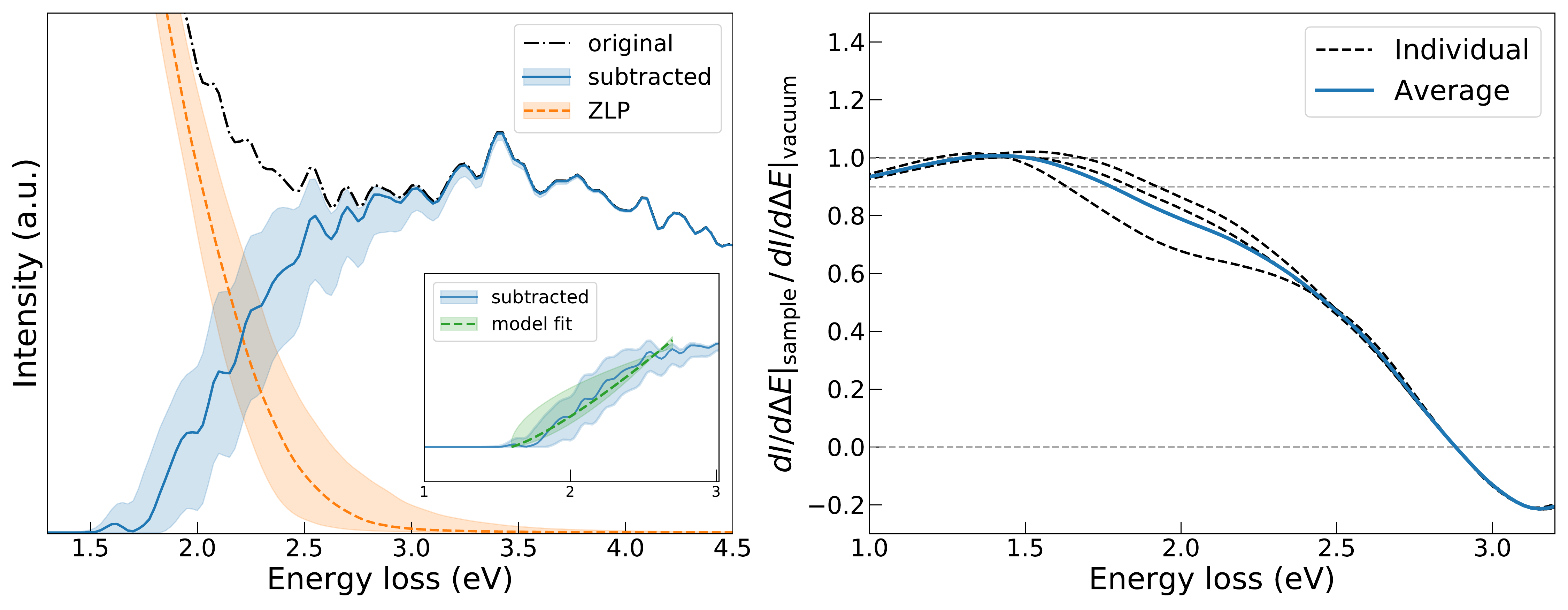}
   \caption{Left: the original
     and subtracted EEL spectra corresponding to location sp4 of sample A in Fig.~\ref{fig:ws2positions},
     together with the predictions of the ZLP model, where
     the bands indicate the 68\% confidence level uncertainties.
     The inset displays the result of fitting Eq.~(\ref{eq:I1}) to the onset
     region of the subtracted spectrum.
     Right: the average ratio of the derivative of the intensity
     distribution in sp4 over its vacuum counterparts, Eq.~(\ref{eq:rder})
  }
\label{fig:sp14_subtracted_spectrum}
\end{centering}
\end{figure}

Here we present the results for the bandgap analysis of sample A,
taking location sp4 in Fig.~\ref{fig:ws2positions} as representative spectrum; compatible
results are obtained for the rest of locations in this sample.
As mentioned above, this region is characterised by a sizable thickness where
WS$_2$ is expected to behave as a bulk material.
The left panel of Fig.~\ref{fig:sp14_subtracted_spectrum} displays the original
and subtracted EEL spectrum
together with the predictions of the ZLP model, where
the bands indicate the 68\% confidence level uncertainties and the central value
is the median of the distribution.
The inset shows the result of the polynomial fits using Eq.~(\ref{eq:I1}) to the subtracted spectrum
together with the corresponding uncertainty bands.

%%%%%%%%%%%%%%%%%%%%%%%%%%%%%%%%%%%%%%%%%%%%%%%%%%%%%%%%%%%%%%%%%%%%%%%
%%%%%%%%%%%%%%%%%%%%%%%%%%%%%%%%%%%%%%%%%%%%%%%%%%%%%%%%%%%%%%%%%%%%%%%%%%

One can observe how the ZLP model uncertainties are small at low $\Delta E$
(due to the matching condition) and large $\Delta E$ (where the ZLP vanishes),
but become significant in the intermediate region where the contributions
from $I_{\rm ZLP}$ and $I_{\rm inel}$ become comparable.
It is worth emphasizing that these (unavoidable) uncertainties are neglected in most
ZLP subtraction methods.
The validity of our choice for the hyperparameter $\Delta E_{\rm I}$ (Table~\ref{table:sampledata_summary})
can be verified {\it a posteriori} by evaluating the ratio
\be
\mathcal{R}^{(j)}_{\rm abs}\lp \Delta E_{\rm I}\rp \equiv 
\la I_{\rm ZLP}^{({\rm mod})(j)}\ra_{\rm rep} \Big/I_{\rm EEL}^{({\rm exp})(j)} \Big|_{\Delta E = \Delta E_{\rm I}} \, ,
\ee
which in this case turns out to be $\mathcal{R}_{\rm abs} = 0.98$.
It is indeed important to verify that $\mathcal{R}_{\rm abs}\lp \Delta E_{\rm I}\rp$ is not too far from unity,
indicating that the training dataset has not been contaminated by the  contributions
arising from inelastic scatterings off the specimen.

\begin{figure}[t]
\begin{centering}
  \includegraphics[width=0.99\linewidth]{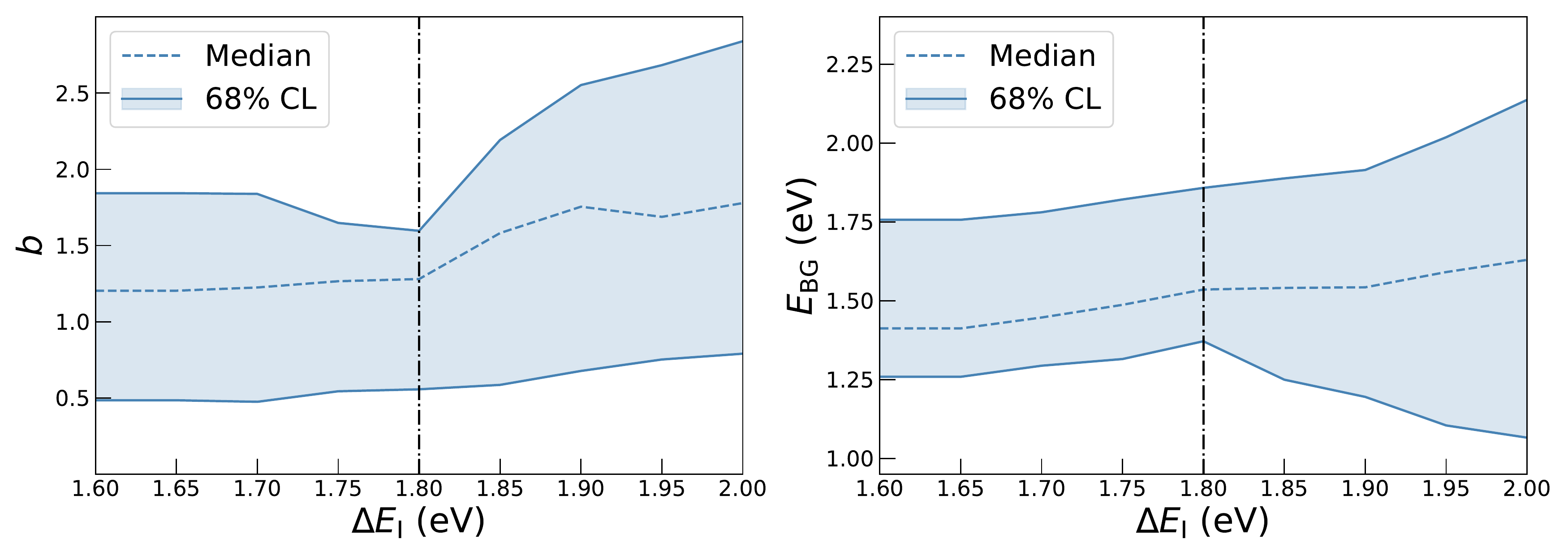} 
  \caption{\small The values of the exponent $b$ (left)
    and the bandgap energy $E_{\rm BG}$ (right panel) from the model Eq.~(\ref{eq:I1})
    obtained from the subtracted spectrum sp14 as $\Delta E_{\rm I}$ is varied by $\pm 0.2$ eV
    around its optimal value, indicated by the horizontal dot-dashed line.
  }
\label{fig:bvalues_sampleA}
\end{centering}
\end{figure}

The average ratio of the derivative of the intensity
distribution in sp4 over its vacuum counterpart, Eq.~(\ref{eq:rder}), is shown
in the right panel of  Fig.~\ref{fig:sp14_subtracted_spectrum}. 
By requiring that $\mathcal{R}_{\rm der}(\Delta E_{\rm I})\simeq 0.9$ we obtain
the value $\Delta E_{\rm I}=1.8$ eV used as baseline in the analysis.
It should be noted that this choice is not unique, for example requiring
$\mathcal{R}_{\rm der}(\Delta E_{\rm I})\simeq 0.8$ instead would have led
to $\Delta E_{\rm I}=2.0$ eV.
It is therefore important to asses the stability of our results as the hyper-parameter $\Delta E_{\rm I}$
is varied around its optimal value.
With this motivation, in Fig.~\ref{fig:bvalues_sampleA} we display the
values of the exponent $b$
and the bandgap energy $E_{\rm BG}$ 
obtained from the same subtracted spectrum as that shown in
Fig.~\ref{fig:sp14_subtracted_spectrum} for variations of $\Delta E_{\rm I}$ 
around its optimal value (vertical dot-dashed line) by an amount
of $\pm 0.2$ eV.
We observe that the model predictions for both $b$ and $E_{\rm BG}$ are stable with respect
to variations of $\Delta E_{\rm I}$, with shifts in central values contained within the
uncertainty bands.
We can thus conclude that our approach is robust with respect to the choice of
hyper-parameters.

%%%%%%%%%%%%%%%%%%%%%%%%%%%%%%%%%%%%%%%%%%%%%%%%%%%%%%%%%%
%%%%%%%%%%%%%%%%%%%%%%%%%%%%%%%%%%%%%%%%%%%%%%%%%%%%%%%%%%%%%

The final values for $E_{\rm BG}$ and $b$ obtained in the analysis of this specific spectrum are
\be
E_{\rm BG} = 1.6_{-0.2}^{+0.3}\,{\rm eV} \, ,\quad b= 1.3_{-0.7}^{+0.3} \, .
\ee
We thus find that for this specific region of the WS$_2$ nanoflowers
the model fit to the subtracted EEL spectrum exhibits a clear preference
for an indirect bandgap (where $b\simeq 1.5$ is expected).
This result is consistent with previous studies of the local
electronic properties of bulk WS$_2$, such as those reported in Table~\ref{table:bgvalues}.
Consistent results are obtained for spectra acquired at
other locations of Fig.~\ref{fig:ws2positions}; for example for sp5 one has
\be
E_{\rm BG} = 1.7 \pm 0.2\,{\rm eV} \, ,\quad b= 1.3_{-0.4}^{+0.3} \, .
\ee
These results represent the first EELS-based bandgap determination of WS$_2$ nanostructures
whose crystalline structure is based on mixed 2H/3R polytypes.

\subsection{Mapping excitonic transitions in the low-loss region}

\begin{figure}[t]
\begin{centering}
  \includegraphics[width=0.99\linewidth]{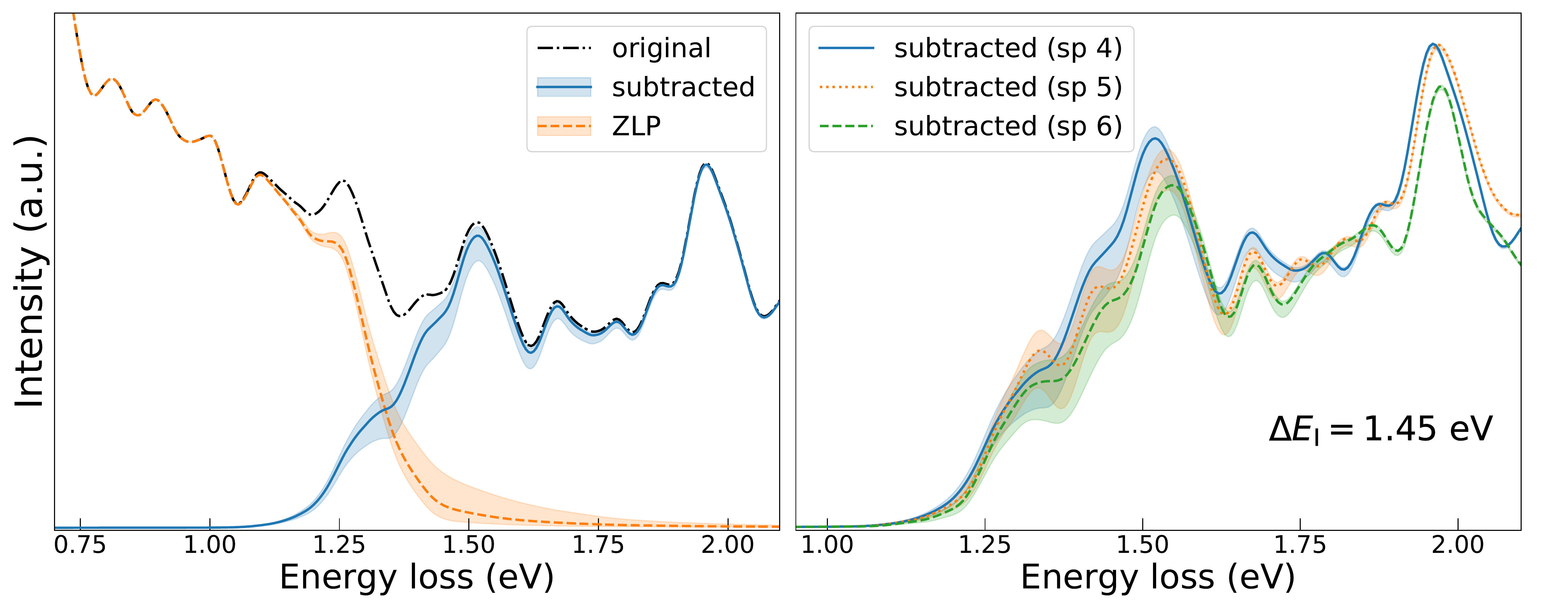}
  \caption{Left: the original
     and subtracted EEL spectra corresponding to location sp4 of sample B in Fig.~\ref{fig:ws2positions},
     together with the predictions of the ZLP model.
     The bands indicate the 68\% confidence level uncertainties.
     Right: comparison of the ZLP-subtracted spectra from locations sp4, sp5, and sp6 in sample B
     together with the corresponding model uncertainties.
     Note how several features of the subtracted spectra, in particular
     the peaks at $\Delta E\simeq 1.5$,
    1.7 and 2.0 are eV, are common across the three locations.
  }
\label{fig:SubtractedEELS_plot_sp4}
\end{centering}
\end{figure}

For the application of our ZLP subtraction strategy to the EEL spectra recorded in specimen B
of the WS$_2$ nanoflowers (bottom panels
in  Fig.~\ref{fig:ws2positions}), the same criterion
based on the derivative ratio Eq.~(\ref{eq:rder}) to select the hyper-parameter $\Delta E_{\rm I}$ was
used.
In this case, one finds a value of $\Delta E_{\rm I}\simeq 1.4$ eV,
 somewhat lower than the corresponding value obtained for sample A.
The left panel of Fig.~\ref{fig:SubtractedEELS_plot_sp4} displays
the original
and subtracted spectra corresponding to the representative
location sp4 of sample B
together with the predictions of the ZLP model.
As before, the bands indicate the 68\% confidence level uncertainties
and the central value is the median.

The main difference with respect to the spectra recorded in sample A is the appearance
of well-defined features (peaks) in the subtracted spectrum already for
very small values of $\Delta E$.
In particular, we observe two marked peaks at $\Delta E\simeq 1.5$ and 2.0 eV and a
softer one near $\Delta E \simeq 1.7$ eV.
Further additional features arise also for higher values of the energy loss.
There are two main sources for the observed differences between the spectra recorded
in each sample.
The first one is that, while sample A is much thicker (bulk material), sample B corresponds
to thin, overlapping petals whose thicknesses can be as small as a few monolayers.
The second is that the EELS measurements taken in sample A used a TEM without monochromator,
while those in sample B benefited from a monochromator thus achieving a
superior
spectral resolution (with an average FWHM of 87 meV to be compared with the 470 meV of sample A, see
Table~\ref{table:sampledata}).
This combination of structural and morphological variations in the specimen together
with the operation conditions of the TEM therefore should account for the
most of differences
between the two sets of spectra.

%%%%%%%%%%%%%%%%%%%%%%%%%%%%%%%%%%%%%%%%%%%%%%%%%%%%%%%%%%%%%%%%%%%%%%%
%%%%%%%%%%%%%%%%%%%%%%%%%%%%%%%%%%%%%%%%%%%%%%%%%%%%%%%%%%%%%%%%%%%%%%%%%%

It is worth noting here that our ZLP parametrisation and subtraction strategy exhibits a satisfactory
performance for all the spectra under consideration, irrespective of the spectral resolution of the TEM used
for their acquisition.
By comparing Figs.~\ref{fig:SubtractedEELS_plot_sp4} and~\ref{fig:sp14_subtracted_spectrum}, one observes
that  model uncertainties are larger in the latter case than in the former, as expected from the
superior 
spectral resolution of the EELS measurements taken on sample B.
Nevertheless, the same approach has been used in both cases without the need of any fine-tuning
or {\it ad hoc} adjustments: of course, if the input
spectra have been recorded with better spectral resolution, the resulting ZLP model uncertainties
will improve accordingly without changing the procedure itself.

Given that the
well-defined spectral features present in Fig.~\ref{fig:SubtractedEELS_plot_sp4}
appear close to the onset of the inelastic emissions, $I_{\rm inel}(\Delta E)$,
these spectra are not suitable for bandgap determination analyses.
The reason is that the method of~\cite{Rafferty:2000}
used in sample A is only applicable under the assumption that there is a sufficiently wide region in $\Delta E$
after the onset of $I_{\rm inel}$ to perform the polynomial fit of Eq.~(\ref{eq:I1}).
This is clearly not possible  for the spectra recorded in sample B, and indeed model fits restricted to $\Delta E\le 1.4$ eV
display a marked numerical instability.
Instead of studying the bandgap properties, it is interesting to exploit the ZLP-subtracted results of sample B
to characterise the local
excitonic transitions of polytypic 2H/3R WS$_2$
that are known to arise in the ultra-low-loss region of the spectra.

Before being able to do this, however, one has to deal with the possible objection
that the peaks present in Fig.~\ref{fig:SubtractedEELS_plot_sp4} are not
genuine features, but rather fluctuations due to insufficient statistics
that should be smoothed out before this region can be interpreted.
To tackle this concern, the right panel of Fig.~\ref{fig:SubtractedEELS_plot_sp4}
displays a
comparison of the ZLP-subtracted spectra recorded in the (spatially separated) locations sp4, sp5 and sp6
in sample B
together with their model uncertainties.
Both the position and the widths of the peaks at $\Delta E\simeq 1.5$,
1.7 and 2.0 eV remain stable, confirming that these
are genuine physical features rather than fluctuations.

These peaks in the ultra-low-loss region of the ZLP-subtracted EELS spectra recorded on thin, polytypic
WS$_2$ nanostructures can be traced back to excitonic transitions.
Their origin can be attributed to the formation of an electron-hole pair mitigated
by the dielectric screening from the surrounding lattice~\cite{Hanbicki:2016}.
In nanostructures with reduced dimensionality as well as in single layers of TMD materials,
exciton peaks arise with binding energies up to ten times larger than for bulk structures.
In the optical spectra of TMDs, two strongly pronounced resonances denoted by A and B
excitons are often observed, appearing at binding energies of 300 and
500 meV below the true bandgap of the material~\cite{Karivaj:2019}.
Interestingly, this prediction is in agreement with the observed peaks at
$\Delta E\simeq 1.5$ and 1.7 eV if one takes into account the expected
value of $E_{\rm BG}$ for very thin  WS$_2$ nanostructures, see Table~\ref{table:bgvalues}

We  conclude that ZLP-subtracted spectra in sample B allow one for
a clean mapping of the exciton peaks present in the WS$_2$ nanoflowers
down to $\Delta E\simeq 1.5$ eV together with
the associated uncertainty estimate.
Further insights concerning the relationship between the exciton peaks in the ultra-low-loss region
and the underlying crystalline structure and specimen morphology could be obtained
by combining our findings with {\it ab initio} calculations such as those based on
density functional theory.

%%%%%%%%%%%%%%%%%%%%%%%%%%%%%%%%%%%%%%%%%
\section{Summary and outlook}
%%%%%%%%%%%%%%%%%%%%%%%%%%%%%%%%%%%%%%%%%
\label{sec:summary}

In this work we have presented a novel, model-independent strategy to parametrise and subtract
the ubiquitous zero-loss peak that dominates  the low-loss region
of EEL spectra.
Our strategy is based on machine learning techniques and provides a faithful estimate of the
uncertainties associated to both the input data and the procedure itself,
which can  then be propagated to physical predictions  without any  approximations.
We have demonstrated how, in the case of vacuum spectra, our approach
is sufficiently flexible to accomodate several input variables corresponding
to different operation conditions of the microscope.
Further, we are able  to reliably
extrapolate our predictions, {\it e.g.} for the  expected FWHM of the ZLP,
to other operation conditions.
When applied to spectra recorded on specimens, our approach
makes possible to robustly disentangle the ZLP contribution from
those arising from inelastic scatterings.
Thanks to this subtraction, one can fully exploit
the valuable physical information contained in the ultra low-loss region of
the spectra.

Here we have applied this ZLP subtraction
strategy to EEL spectra recorded in  WS$_2$ nanoflowers characterised by a
2H/3R polytypic crystalline structure.
First of all, measurements taken in a relatively
thick region of the specimen were used to determine
the local value of the bandgap energy $E_{\rm BG}$
and to assess whether this bandgap is direct or indirect.
A model fit to the onset of the inelastic intensity distribution obtains
$E_{\rm BG} \simeq 1.6^{+0.3}_{-0.2}\,{\rm eV}$ and exhibits a marked preference for an indirect bandgap.
Our findings are consistent with previous studies, both of theoretical
and of experimental nature, concerning the bandgap structure of bulk WS$_2$.

Subsequently, we have applied our method to a  thinner region of the  WS$_2$ nanoflowers,
specifically a region composed by overlapping petals with varying
thicknesses that can be as small as a few monolayers.
We have demonstrated how for such specimens one can exploit the ZLP-subtracted results
to characterise the local excitonic transitions that arise in the ultra-low-loss region.
By charting the exciton peaks of 2H/3R polytypic WS$_2$ there,
we identify two strong peaks at $\Delta E\simeq 1.5$ and 2 eV
(and a softer one at 1.7 eV) and show how
these features are consistent when comparing
spatially-separated locations in sample B.
Further, since our method provides an associated uncertainty estimate,
one can robustly establish the
statistical significance of each of these
ultra-low-loss region features.

The approach presented in this work could be extended
in several directions.
First of all, it would be interesting to test its robustness when additional
operation conditions of the microscope are included as input variables,
and to verify to which extent the ZLP parametrisations obtained for an specific microscope
can be generalised to an altogether different TEM.
Further, a non-trivial cross-check of our method would be provided by validating
our predictions for other operation conditions of the microscope, such
as the FWHM as a function of the beam energy $E_b$ of the exposure time
$t_{\rm exp}$ reported in Fig.~\ref{fig:extrapolbeam},
with actual measurements.

Concerning the physical interpretation of the low-loss region of EEL
spectra, our method could be applied to study the bandgap properties 
for different types
of nanostructures built upon TMD materials, such as MoS$_2$ nanowalls~\cite{nanowalls}
and 
vertically-oriented nano-sheets~\cite{D0NR00755B} or
WS$_2$/MoS$_2$ arrays, heterostructures, and ternary alloys.
In addition to bandgap characterisation, this ZLP-subtraction
strategy should allow the detailed study
of other phenomena relevant for the interpretation of the low-loss
region such as  plasmons, excitons, phonon interactions, and
intra-band transitions.
One could also exploit the subtracted EEL spectra to further characterise
local electronic properties by means of the
 evaluation of the complex dielectric function and its associated
uncertainties in terms of the Kramers-Kronig relations.
Finally, these phenomenological studies of local electronic properties should be compared
with {\it ab initio} calculations based
on the same underlying crystalline structure as the studied specimens.

Another possible application of the strategy presented in this work would be the automation of
the study of spectral TEM images,
such as those displayed in the right panels of Fig.~\ref{fig:ws2positions},
where each pixel contains an individual EEL spectrum.
Here machine learning methods would provide a  useful handle in order
to identify relevant features of the spectra (peaks, edges, shoulders) with minimal
human intervention (no need to process each spectrum individually) and then determine
how these features vary as we move along different regions of the
nanostructure.
Such an approach would combine two important families of machine learning algorithms,
those used for regression, in order to quantify the properties of spectral
features such as width and significance, and those for classification, to identify categories
of distinct features across the spectral image.

\subsection*{Acknowledgments}
We are grateful to Emanuele R. Nocera and Jacob J. Ethier for
assistance in installing {\tt EELSfitter} in the Nikhef computing cluster.
L.~R. is grateful to Cas, Agneet, and Aar, for support under all 
(rainy) circumstances.

\subsection*{Funding}

S.~E.~v.~H. and S.~C.-B. acknowledge financial support
from the ERC through the Starting Grant ``TESLA”'', grant agreement
no. 805021.
L.~M. acknowledges support from the
Netherlands Organizational for Scientific Research (NWO)
through the Nanofront program.
The work of J.~R. has been partially supported by NWO.

\subsection*{Declaration of competing interest}

The authors declare that they have no known competing financial interests or personal relationships that could have appeared to influence the work reported in this paper.

\subsection*{Methods}

{\justify
The EEL spectra used for the training of the vacuum ZLP model presented in Sect.~\ref{sec:results_vacuum} were collected in a ARM200F Mono-JEOL microscope equipped with a GIF continuum spectrometer and operated at 60 kV and 200 kV. For these measurements, a slit in the monochromator of 2.8 $\mu$m was used.
The TEM and EELS measurements acquired in Specimen A for the results presented in
Sect.~\ref{sec:results_sample} were recorded in a JEOL 2100F microscope with a cold field-emission
gun equipped with aberration corrector operated at 60 kV. A Gatan GIF Quantum was used for
the EELS analyses. The convergence and collection semi-angles were 30.0 mrad and 66.7 mrad respectively.
The TEM and EELS measurements acquired for Specimen B in Sect.~\ref{sec:results_sample}
were recorded using a JEM ARM200F monochromated microscope operated at 60 kV and equipped with
a GIF quantum ERS. The convergence and collection semi-angles were 24.6 mrad and 58.4 mrad respectively
in this case, and the aperture of the spectrometer was set to 5 mm.}

%\bibliography{bib/EELS_ML}

\begin{thebibliography}{10}

\bibitem{Terauchi:2005}
M.~Terauchi, T.~M., K.~Tsuno, and M.~Ishida, {\it {Development of a high energy
  resolution electron energy-loss spectroscopy microscope}},  {\em Journal of
  Microscopy} {\bf 194} (2005) 203--209.

\bibitem{Freitag:2005}
B.~Freitag, S.~Kujawa, P.~Mul, J.~Ringnalda, and P.~Tiemeijer, {\it {Breaking
  the spherical and chromatic aberration barrier in transmission electron
  microscopy}},  {\em Ultramicroscopy} {\bf 102} (2005) 209--214.

\bibitem{Haider:1998}
M.~Haider, S.~Uhlemann, E.~Schwan, H.~Rose, B.~Kabius, and K.~Urban, {\it
  {Electron microscopy image enhanced}},  {\em Nature} {\bf 392} (1998)
  768–769.

\bibitem{Geiger:1967}
J.~Geiger, {\it {Inelastic Electron Scattering in Thin Films at Oblique
  Incidence}},  {\em Phys. Stat. Sol.} {\bf 24} (1967) 457--460.

\bibitem{Schaffer:2008}
B.~Schaffer, K.~Riegler, G.~Kothleitner, W.~Grogger, and F.~Hofer, {\it
  {Monochromated, spatially resolved electron energy-loss spectroscopic
  measurements of gold nanoparticles in the plasmon range}},  {\em Micron} {\bf
  40} (2008) 269--273.

\bibitem{Erni:2005}
R.~Erni, N.~D. Browning, Z.~Rong~Dai, and J.~P. Bradley, {\it {Analysis of
  extraterrestrial particles using monochromated electron energy-loss
  spectroscopy}},  {\em Micron} {\bf 35} (2005) 369–379.

\bibitem{Rafferty:1998}
B.~Rafferty and L.~Brown, {\it {Direct and indirect transitions in the region
  of the band gap using electron-energy-loss spectroscopy}},  {\em Physical
  Review B} {\bf 58} (1998) 10326.

\bibitem{Stoger:2008}
M.~Stöger-Pollach, {\it {Optical properties and bandgaps from low loss EELS:
  Pitfalls and solutions}},  {\em Nano Lett.} {\bf 39} (2008) 1092–1110.

\bibitem{Egerton:2009}
R.~Egerton, {\it {Electron energy-loss spectroscopy in the TEM}},  {\em Rep.
  Prog. Phys} {\bf 72} (2009) 1.

\bibitem{Abajo:2010}
F.~García~de Abajo, {\it {Optical excitations in electron microscopy}},  {\em
  RevModPhys} {\bf 82} (2010) 209--256,
  [\href{http://arxiv.org/abs/0903.1669}{{\tt arXiv:0903.1669}}].

\bibitem{Park:2008}
J.~Park, S.~Heo, et~al., {\it {Bandgap measurement of thin dielectric films
  using monochromated STEM-EELS}},  {\em Ultramicroscopy} {\bf 109} (2008)
  1183–1188.

\bibitem{Rafferty:2000}
B.~Rafferty, S.~Pennycook, and L.~Brown, {\it {Zero loss peak deconvolution for
  bandgap EEL spectra}},  {\em Journal of Electron Microscopy} {\bf 49} (2000)
  517--524.

\bibitem{Egerton:1996}
R.~Egerton, {\em {Electron Energy-Loss Spectroscopy in the Electron
  Microscope}}.
\newblock Plenum Press, 1996.

\bibitem{Dorneich:1998}
A.~Dorneich, R.~French, H.~Müllejans, et~al., {\it {Quantitative analysis of
  valence electron energy-loss spectra of aluminium nitride}},  {\em Journal of
  Microscopy} {\bf 191} (1998) 286–296.

\bibitem{Benthem:2001}
K.~van Benthem, C.~Elsässer, and R.~French, {\it {Bulk electronic structure of
  SrTiO3: Experiment and theory}},  {\em Journal of Applied Physics} {\bf 90}
  (2001).

\bibitem{Lazar:2003}
S.~Lazar, G.~Botton, et~al., {\it {Materials science applications of HREELS in
  near edgestructure analysis an low-energy loss spectroscopy}},  {\em
  Ultramicroscopy} {\bf 96} (2003) 535–546.

\bibitem{Egerton:10.1016/S0304-3991(01)00155-3}
R.~Egerton and M.~Malac, {\it {Improved background-fitting algorithms for
  ionization edges in electron energy-loss spectra}},  {\em Ultramicroscopy}
  {\bf 92} (2002) 47--56.

\bibitem{Held:2020}
J.~T. Held, H.~Yun, and K.~A. Mkhoyan, {\it {Simultaneous multi-region
  background subtraction for core-level EEL spectra}},  {\em Ultramicroscopy}
  {\bf 210} (2020) 112919.

\bibitem{Granerod:2018}
C.~S. Granerød, W.~Zhan, and Øystein Prytz, {\it {Automated approaches for
  band gap mapping in STEM-EELS}},  {\em Ultramicroscopy} {\bf 184} (2018)
  39--45.

\bibitem{Fung:2020}
K.~L. Fung, M.~W. Fay, S.~M. Collins, D.~M. Kepaptsoglou, S.~T. Skowron, Q.~M.
  Ramasse, and A.~N. Khlobystov, {\it {Accurate EELS background subtraction, an
  adaptable method in MATLAB}},  {\em Ultramicroscopy} {\bf 217} (2020) 113052.

\bibitem{Ball:2008by}
{\bf The NNPDF} Collaboration, R.~D. Ball et~al., {\it {A determination of
  parton distributions with faithful uncertainty estimation}},  {\em Nucl.
  Phys.} {\bf B809} (2009) 1--63, [\href{http://arxiv.org/abs/0808.1231}{{\tt
  arXiv:0808.1231}}].

\bibitem{Ball:2012cx}
R.~D. Ball et~al., {\it {Parton distributions with LHC data}},  {\em
  Nucl.Phys.} {\bf B867} (2013) 244,
  [\href{http://arxiv.org/abs/1207.1303}{{\tt arXiv:1207.1303}}].

\bibitem{Ball:2014uwa}
{\bf NNPDF} Collaboration, R.~D. Ball et~al., {\it {Parton distributions for
  the LHC Run II}},  {\em JHEP} {\bf 04} (2015) 040,
  [\href{http://arxiv.org/abs/1410.8849}{{\tt arXiv:1410.8849}}].

\bibitem{Ball:2017nwa}
{\bf NNPDF} Collaboration, R.~D. Ball et~al., {\it {Parton distributions from
  high-precision collider data}},  {\em Eur. Phys. J.} {\bf C77} (2017), no.~10
  663, [\href{http://arxiv.org/abs/1706.00428}{{\tt arXiv:1706.00428}}].

\bibitem{SabryaWS2}
  S.~E. van Heijst, M.~Masaki, E.~Okunishi, H.~Hashiguchi,
  L.~I. Roest, L.~Maduro, J.~Rojo, and S.~Conesa-Boj,
  {\it {Fingerprinting 2H/3R Polytypism in WS$_2$
  Nanoflowers from Plasmons and Excitons to Phonons}} (2020), in preparation.

\bibitem{Bangert:2003}
U.~Bangert, A.~Harvey, D.~Freundt, and R.~Keyse, {\it {Highly spatially
  resolved electron energy‐loss spectroscopy in the bandgap regime of GaN}},
  {\em Journal of Microscopy} {\bf 188} (1998) 237--242.

\bibitem{Hachtel:2018}
J.~Hachtel, A.~Lupini, and J.~Idrobo, {\it {Exploring the capabilities of
  monochromated electron energy loss spectroscopy in the infrared regime}},
  {\em Scientific Reports} {\bf 8} (2018) 5637.

\bibitem{Tenailleau:1992}
H.~Tenailleau and J.~M. Martin, {\it {A new background subtraction for
  low‐energy EELS core edges}},  {\em Journal of Microscopy} {\bf 116} (1992)
  297–306.

\bibitem{Reed:2002}
B.~Reed and M.~Sarikaya, {\it {Background subtraction for low-loss transmission
  electron energy-loss spectroscopy}},  {\em Ultramicroscopy} {\bf 93} (2002)
  25--37.

\bibitem{Bosman:2006}
M.~Bosman, M.~Watanabe, D.~Alexander, and V.~Keast, {\it {Mapping chemical and
  bonding information using multivariate analysis of electron energy-loss
  spectrum images}},  {\em Ultramicroscopy} {\bf 106} (2006) 1024--1032.

\bibitem{Kothleitner:2003}
G.~Kothleitner and F.~Hofer, {\it {EELS performance measurements on a new high
  energy resolution imaging filter}},  {\em Micron} {\bf 34} (2003) 211--218.

\bibitem{Chhowalla:2013}
M.~Chhowalla, H.~S. Shin, G.~Eda, L.-J. Li, K.~P. Loh, and H.~Zhangs, {\it {The
  chemistry of two-dimensional layered transition metal dichalcogenide
  nanosheets}},  {\em Nature Chemistry} {\bf 5} (2013) 263--275.

\bibitem{Splendiani:2010}
A.~Splendiani, L.~Sun, T.~Li, et~al., {\it {Emerging Photoluminescence in
  Monolayer MoS$_2$}},  {\em Nano Lett.} {\bf 10} (2010) 1271–1275.

\bibitem{Zhao2013}
W.~Zhao, Z.~Ghorannevis, L.~Chu, M.~Toh, C.~Kloc, P.-H. Tan, and G.~Eda, {\it
  Evolution of electronic structure in atomically thin sheets of WS$_2$ and WSe$_2$},
   {\em ACS Nano} {\bf 7} (01, 2013) 791--797.

\bibitem{Bhandavat:2012}
R.~Bhandavat, L.~David, and G.~Singh, {\it {Synthesis of Surface-Functionalized
  WS$_2$ Nanosheets and Performance as Li-Ion Battery Anodes}},  {\em J. Phys.
  Chem. Lett.} {\bf 3} (2012) 1523–1530.

\bibitem{Na:2018}
W.~Na, K.~Kim, J.-U. Lee, and H.~Cheong, {\it {Davydov splitting and polytypism
  in few-layer MoS$_2$}},  {\em 2D Materials} {\bf 6} (2018) 015004.

\bibitem{Lee:2016}
J.-U. Lee, K.~Kim, S.~Han†, G.~H. Ryu, Z.~Lee, and H.~Cheong, {\it {Raman
  Signatures of Polytypism in Molybdenum Disulfide}},  {\em ACS Nano} {\bf 10}
  (2016) 1948–1953.

\bibitem{XIA20171}
J.~Xia, J.~Yan, and Z.~X. Shen, {\it Transition metal dichalcogenides:
  structural, optical and electronic property tuning via thickness and
  stacking},  {\em FlatChem} {\bf 4} (2017) 1 -- 19.

\bibitem{Braga:2012}
D.~Braga, I.~G. Lezama, H.~Berger, and A.~F. Morpurgo, {\it {Quantitative
  Determination of the Band Gap of WS$_2$ with Ambipolar Ionic Liquid-Gated
  Transistors}},  {\em NanoLetters} {\bf 12} (2012) 5218.

\bibitem{Jo:2014}
S.~Jo, N.~Ubrig†, H.~Berger, A.~B. Kuzmenko†, and A.~F. Morpurgo, {\it
  {Mono- and Bilayer WS2 Light-Emitting Transistors}},  {\em Nano Letters} {\bf
  14} (2014) 2019--2025.

\bibitem{Gusakova:2007}
J.~G. et~al, {\it {Electronic Properties of Bulk and Monolayer TMDs:
  Theoretical Study Within DFT Framework (GVJ‐2e Method)}},  {\em Physica
  Status Solidi A} {\bf 214} (2007).

\bibitem{Kam:1982}
K.~Kam and B.~Parkinson, {\it {Detailed photocurrent spectroscopy of the
  semiconducting group VIB transition metal dichalcogenides}},  {\em J. Phys.
  Chem.} {\bf 86} (1982) 463–467.

\bibitem{Shi:2013}
H.~Shi, H.~Pan, Y.-W. Zhang, and B.~I. Yakobson, {\it {Quasiparticle band
  structures and optical properties of strained monolayer MoS$_2$ and WS$_2$}},  {\em
  PHYSICAL REVIEW B} {\bf 87} (2013) 155304.

\bibitem{Rojo:2018qdd}
J.~Rojo, {\it {Machine Learning tools for global PDF fits}},  in {\em {13th
  Conference on Quark Confinement and the Hadron Spectrum}}, 9, 2018.
\newblock \href{http://arxiv.org/abs/1809.04392}{{\tt arXiv:1809.04392}}.

\bibitem{Gao:2017yyd}
J.~Gao, L.~Harland-Lang, and J.~Rojo, {\it {The Structure of the Proton in the
  LHC Precision Era}},  {\em Phys. Rept.} {\bf 742} (2018) 1--121,
  [\href{http://arxiv.org/abs/1709.04922}{{\tt arXiv:1709.04922}}].

\bibitem{DelDebbio:2007ee}
{\bf The NNPDF} Collaboration, L.~Del~Debbio, S.~Forte, J.~I. Latorre,
  A.~Piccione, and J.~Rojo, {\it {Neural network determination of parton
  distributions: The nonsinglet case}},  {\em JHEP} {\bf 03} (2007) 039,
  [\href{http://arxiv.org/abs/hep-ph/0701127}{{\tt hep-ph/0701127}}].

\bibitem{Nocera:2014gqa}
{\bf NNPDF Collaboration} Collaboration, E.~R. Nocera, R.~D. Ball, S.~Forte,
  G.~Ridolfi, and J.~Rojo, {\it {A first unbiased global determination of
  polarized PDFs and their uncertainties}},  {\em Nucl.Phys.} {\bf B887} (2014)
  276, [\href{http://arxiv.org/abs/1406.5539}{{\tt arXiv:1406.5539}}].

\bibitem{AbdulKhalek:2019mzd}
{\bf NNPDF} Collaboration, R.~Abdul~Khalek, J.~J. Ethier, and J.~Rojo, {\it
  {Nuclear parton distributions from lepton-nucleus scattering and the impact
  of an electron-ion collider}},  {\em Eur. Phys. J.} {\bf C79} (2019), no.~6
  471, [\href{http://arxiv.org/abs/1904.00018}{{\tt arXiv:1904.00018}}].

\bibitem{AbdulKhalek:2020yuc}
R.~Abdul~Khalek, J.~J. Ethier, J.~Rojo, and G.~van Weelden, {\it {nNNPDF2.0:
  Quark Flavor Separation in Nuclei from LHC Data}},
  \href{http://arxiv.org/abs/2006.14629}{{\tt arXiv:2006.14629}}.

\bibitem{Bertone:2017tyb}
{\bf NNPDF} Collaboration, V.~Bertone, S.~Carrazza, N.~P. Hartland, E.~R.
  Nocera, and J.~Rojo, {\it {A determination of the fragmentation functions of
  pions, kaons, and protons with faithful uncertainties}},  {\em Eur. Phys. J.
  C} {\bf 77} (2017), no.~8 516, [\href{http://arxiv.org/abs/1706.07049}{{\tt
  arXiv:1706.07049}}].

\bibitem{Bertone:2018ecm}
{\bf NNPDF} Collaboration, V.~Bertone, N.~Hartland, E.~Nocera, J.~Rojo, and
  L.~Rottoli, {\it {Charged hadron fragmentation functions from collider
  data}},  {\em Eur. Phys. J. C} {\bf 78} (2018), no.~8 651,
  [\href{http://arxiv.org/abs/1807.03310}{{\tt arXiv:1807.03310}}].

\bibitem{Gordon:2020}
O.~M. Gordon and P.~J. Moriarty, {\it {Machine learning at the (sub)atomic
  scale: next generation scanning probe microscopy}},  {\em Machine Learning:
  Science and Technology} {\bf 1} (2020) 023001.

\bibitem{Zhang:2019}
Y.~Zhang, A.~Mesaros, and K.~e. Fujita, {\it {Machine learning in
  electronic-quantum-matter imaging experiments}},  {\em Nature} {\bf 570}
  (2020) 484–490.

\bibitem{Jany:2017}
B.~Jany, A.~Janas, and F.~Krok, {\it {Retrieving the Quantitative Chemical
  Information at Nanoscale from Scanning Electron Microscope Energy Dispersive
  X-ray Measurements by Machine Learning}},  {\em Nano Letters} {\bf 17} (2017)
  6520–6525.

\bibitem{Ziatdinov:2017}
M.~Ziatdinov, O.~Dyck, A.~Maksov, X.~Li, X.~Sang, K.~Xiao, R.~R. Unocic,
  R.~Vasudevan, S.~Jesse, and S.~V. Kalinin, {\it {Deep Learning of Atomically
  Resolved Scanning Transmission Electron Microscopy Images: Chemical
  Identification and Tracking Local Transformations}},  {\em ACS Nano} {\bf 11}
  (2017) 12742--12752.

\bibitem{10.1145/2834892.2834896}
S.~R. Young, D.~C. Rose, T.~P. Karnowski, S.-H. Lim, and R.~M. Patton, {\it
  Optimizing deep learning hyper-parameters through an evolutionary algorithm},
   in {\em Proceedings of the Workshop on Machine Learning in High-Performance
  Computing Environments}, MLHPC ’15, (New York, NY, USA), Association for
  Computing Machinery, 2015.

\bibitem{doi:10.1021/acsnano.7b07504}
M.~Ziatdinov, O.~Dyck, A.~Maksov, X.~Li, X.~Sang, K.~Xiao, R.~R. Unocic,
  R.~Vasudevan, S.~Jesse, and S.~V. Kalinin, {\it Deep learning of atomically
  resolved scanning transmission electron microscopy images: Chemical
  identification and tracking local transformations},  {\em ACS Nano} {\bf 11}
  (2017), no.~12 12742--12752,
  [\href{http://arxiv.org/abs/https://doi.org/10.1021/acsnano.7b07504}{{\tt
  https://doi.org/10.1021/acsnano.7b07504}}]. PMID: 29215876.

\bibitem{cite-key}
K.~de~Haan, Z.~S. Ballard, Y.~Rivenson, Y.~Wu, and A.~Ozcan, {\it Resolution
  enhancement in scanning electron microscopy using deep learning},  {\em
  Scientific Reports} {\bf 9} (2019), no.~1 12050.

\bibitem{Forte:2002fg}
S.~Forte, L.~Garrido, J.~I. Latorre, and A.~Piccione, {\it Neural network
  parametrization of deep-inelastic structure functions},  {\em JHEP} {\bf 05}
  (2002) 062, [\href{http://arxiv.org/abs/hep-ph/0204232}{{\tt
  hep-ph/0204232}}].

\bibitem{Abadi:2016kic}
M.~Abadi et~al., {\it {TensorFlow: Large-Scale Machine Learning on
  Heterogeneous Distributed Systems}},
  \href{http://arxiv.org/abs/1603.04467}{{\tt arXiv:1603.04467}}.

\bibitem{Schamm:2003}
S.~Schamm and G.~Zanchi, {\it {Study of the dielectric properties near the band
  gap by VEELS:gap measurement in bulk materials}},  {\em Ultramicroscopy} {\bf
  96} (2003) 559--564.

\bibitem{Hanbicki:2016}
A.~Hanbicki, M.Currie, G.Kioseoglou, A.L.Friedman, and B.T.Jonker, {\it
  {Measurement of high exciton binding energy in the monolayer transition-metal
  dichalcogenides WS$_2$ and WSe$_2$}},  {\em Solid State Communications} {\bf 204}
  (2016) 16--20.

\bibitem{Karivaj:2019}
B.~Kaviraj and D.~Sahoo, {\it {Physics of excitons and their transport in two
  dimensional transition metal dichalcogenide semiconductors}},  {\em RSC
  Advances} (2019).

\bibitem{nanowalls}
M.~Tinoco, L.~Maduro, and S.~Conesa-Boj, {\it Metallic edge states in zig-zag
  vertically-oriented MoS$_2$ nanowalls},  {\em Scientific Reports} {\bf 9}
  (2019), no.~1 15602.

\bibitem{D0NR00755B}
M.~Bolhuis, J.~Hernandez-Rueda, S.~E. van Heijst, M.~Tinoco~Rivas, L.~Kuipers,
  and S.~Conesa-Boj, {\it Vertically-oriented MoS$_2$ nanosheets for nonlinear
  optical devices},  {\em Nanoscale} {\bf 12} (2020) 10491--10497.

\end{thebibliography}
\providecommand{\href}[2]{#2}\begingroup\raggedright\endgroup

\appendix
%\vspace{1cm}
%\hrule
%\vspace{1cm}

\clearpage
\begin{center}
  {\bf \LARGE Supplementary Material}
 \end{center}

\section{Installation and usage of {\tt EELSfitter}}
\label{sec:installation}

In this appendix we provide some instructions about the installation
and the usage of the {\tt EELSfitter} code developed
in this work.
The code is available from its GitHub repository
\begin{center}
\url{https://github.com/LHCfitNikhef/EELSfitter}
\end{center}
and is composed by a number of {\tt Python} scripts.
The code requires a working installation of {\tt Python3} and the following
libraries: {\tt NumPy}, {\tt TensorFlow} (v2), {\tt pandas}, {\tt SciPy} and {\tt scikit-learn}.

\noindent
\paragraph{\tt Load\_data.py}
This script reads the spectrum
intensities and create data-frames to be used for training the neural network.
It reads out the EEL spectra intensities, automatically selects the energy loss
at which the peak intensity occurs and shifts the dataset such that
the peak intensity is centered at $\Delta E =$0. 
Further, for each spectrum it returns the normalized intensity by normalizing
over the total area under the spectrum. 
The output is two datasets, {\tt df} and {\tt df\_vacuum} which contain the 
information on the in-sample and in-vacuum recorded spectra respectively. 
The user needs to upload the spectral data in .txt format to the 'Data' folder
and make sure that the vacuum and in-sample spectra are added to the appropriate one.
For each of the spectra the minimum and maximum value of the recorded energy 
loss need to be set manually in {\tt Eloss\_min} and {\tt Eloss\_max}.

\noindent
\paragraph{\tt Fitter.ipynb}
This script is used to run the neural network training on the data that was 
uploaded using {\tt load\_data.py}.
It involves a number of pre-processing steps to determine the hyper-parameters $\Delta E_{\rm I}$
and $\Delta E_{\rm II}$ and then
it automatically prepares and cuts the data before it is fed
to the neural network to start the training.
It is structured as follows:

\begin{itemize}
  
\item {\it Importing libraries and spectral data} from the {\bf load\_data.py} script.

\item {\it Evaluate  $\Delta E_{\rm I}$ from the intensity derivatives}.
  In order to determine the value for the hyper-parameter
  $\Delta E_{\rm I}$, a dataframe {\tt df\_dx} is created
and it calculates the derivatives of each of the in-sample recorded spectra, 
stored as {\tt df\_dx['derivative y*']}, where {\tt *} is any of the in-sample recorded spectra.
The first crossing of any of the derivatives with zero is determined 
and stored as the value of $\Delta E_{\rm I}$. 

\item {\it Evaluate $\Delta E_{\rm II}$ for the pseudo-data.}
It calculates the mean over all vacuum spectra, {\tt df\_mean}, and the ratio of the 
intensity to the experimental uncertainty for each value of
$\Delta E$, {\tt df\_mean['ratio']}. 
The value of $\Delta E_{\rm II}$ is then determined as the energy loss at which this ratio
drops below 1 and  is stored together with the value of $\Delta E_{\rm I}$
as the hyper-parameters for training. 
However, if one wishes to use other values for these parameters, for instance for 
cross-validating the best value for $\Delta E_{\rm I}$, these can also be adjusted manually.

\item {\it Experimental data processing.}
  The next step is to keep only
  the data points with $\Delta E \le \Delta E_{\rm I}$  
and dropping the points with higher energy losses.
Experimental central values and uncertainties are calculated by means of equal width 
discretization, for which the number of bins has to be set as {\tt nbins}. 
The default value is 32, which means that 32 training inputs are spread equally
over the range [$ \Delta E_{\rm min}, \Delta E_{\rm I}$]. 
Note that the logarithm of the intensity is used as training inputs, because this facilitates
the optimization of the neural network ($I_{\rm EEL}$ being a steeply falling function
of $\Delta E$).
The code translates this back to the original intensity values after the training.
$N_{\rm pd}$
pseudo datapoints are added in the range $[\Delta E_{\rm II}, \Delta E_{\rm max}]$, where
$ \Delta E_{\rm max}$
is the maximum energy loss value of the recorded spectra. 
The values for $N_{\rm pd}$ and $\Delta E_{\rm max}$ should be changed manually by 
setting them in {\tt max\_x} and {\tt N\_pseudo}. 
The output is a dataframe {\tt df\_full} containing all training data and pseudo data points, 
corresponding to a total of $N_{\rm in}$ (= $n_{\rm bins}$ + $N_{\rm pd}$) training inputs.

\item {\it Initialize the NN model,} where the code
  defines the neural network architecture and prepares the
data inputs to feed them to the neural network for training. 
The function {\tt make\_model()} allows to define the number of hidden layers and 
nodes per layer. The default  architecture is 1-10-15-5-1.

\item {\it Initialize data for NN training.}
  Here the code prepares the recorded spectra to be used
as inputs for the neural network. 
First, we initiate placeholders for the variables
{\tt x}, {\tt y} and {\tt sigma} which
allow us to create our operations and 
build our computation graph, without needing the data itself. 
The dimension of the placeholder is defined by {\tt $[{\bf None}, dim]$} where 'dim'
should be set to the dimension of the corresponding variable. In this case the input is 
one-dimensional, so dim=1. 
These placeholders are used to define {\tt predictions}, which is in fact a placeholder that is used 
later to make predictions on inputs {\tt x}. 
Also, we define a vector {\tt predict\_x} that is used to make a direct prediction after training
on each of the replicas. It consists of $N_{\rm pred}$ data points in the energy loss range
{\tt [pred\_min, pred\_max]}. 

\item {\it Create the Monte Carlo replicas.}
  The final step to be taken before we can start training is the creation of
  sample of $N_{\rm rep}$ Monte Carlo replicas of the original EEL spectra,
  following the procedure described in Sect.~\ref{sec:MCreplicas}.
This is done automatically using the experimental intensities {\tt train\_y} and uncertainties
{\tt train\_sigma} for a total of {\tt Nrep} replicas. The output is an ($N_{\rm in}, N_{\rm rep}$) 
vector containing all the MC replicas. 

\item {\it Train the neural networks.}
  The final part of the script, where the NN training is  carried out,
  is based on the function {\tt function\_train()} that
  implements the strategy presented in Sect.~\ref{sec:training}.
The cost function, optimizer and learning rate are defined here, together with a 'saver' used to 
save the network parameters after each optimization. 
We start a loop over {\tt Nrep} replicas to initiate a training session on each of the individual replicas
in series. 
For each iteration, the $k$-th replica is selected from the sample of $N_{\rm rep}$ replicas.
The data is split into 80\% training and 20\% validation data, this partition is done 
at random for each replica. The resulting {\tt train\_y} and {\tt test\_y} arrays are used
as training and validation labels.
The total number of training epochs per session is defined in {\tt training\_epochs}.
The script displays intermediate 
results after each number of epochs defined by {\tt display\_step}. 
Running the session object over the optimizer and cost function requires knowledge about the values of {\tt x} and {\tt sigma}, which 
are defined inside the {\tt feed\_dict} argument. 
After each epoch the average training  validation costs are evaluated
and the network parameters  updated accordingly.

Once the maximum number of epochs 
had been reached, the optimal stopping point is determined by 
taking the absolute minimum of the validation cost
and restoring the corresponding network parameters by means of the 'saver' function.
From this network graph, one can directly output the prediction on the values of {\tt train\_x} and
the results are stored in the array {\tt predictions\_values}.
It is also possible to make predictions on any input vector of choice by feeding 
the vector {\tt predict\_x} to the 
network, which outputs an array {\tt extrapolation}.

\end{itemize}

The datafiles that are stored upon successfully
executing this script are the following:

\begin{itemize}

\item {\tt Prediction\_k} contains the energy loss {\tt train\_x}, the MC training data {\tt train\_y}
and the ZLP prediction made on the array {\tt train\_x}, where {\tt k} is the $k$-th replica. 
\item {\tt Cost\_k} contains the training and validation error for the
  $k$-th replica
stored after each display step. 
The minimum of the the validation array is used to restore the optimal
neural network parameters.
\item {\tt Extrapolation\_k} contains the arrays {\tt predict\_x} and the ZLP predictions made on these values. 
\end{itemize}
These text files can be retrieved later to make new ZLP predictions
without the need to repeat the training procedure.
Futher, we store the optimal network parameters after each training session in the folder
'Models/Best\_models'. 
These can be loaded at a later stage
to make predictions for an arbitrary set of input variables. 

Running the loop over all replicas in series, using an input array of $\sim$50 training points 
and a total number of training epochs of 25000 per session,
takes approximately 20 seconds per optimization ($\sim$200 replicas per hour).

\noindent
\paragraph{\tt predictions.ipynb}
This script is used to analyse the predictions from the trained
neural networks that have been stored in the text files indicated above.

\begin{itemize}
  
\item {\it Import libraries and spectral data} from the {\bf load\_data.py} script.

\item {\it Create dataframes with all individual spectra.}
In order to later subtract all the predictions from the original individual spectra, we create a datafile
{\tt original} which contains the intensity values for each of the original input spectra restricted to the region between
 {\tt E\_min} and {\tt E\_max}.

\item {\it Load result files.}
In order to import the files that were stored during the NN training, 
one should input to this script the right directions to find the prediction .txt files
by adjusting the lines {\tt path\_to\_data} and {\tt path\_predict}, {\tt path\_cost} and {\tt path\_extrapolate}, 
containing the predictions, cost function data and the extrapolation predictions respectively.

\item {\it Post-selection criteria.}
  Here one select the datafiles that satisfy suitable post-fit selection
  criteria, such as the final error function being smaller
  than a certain threshold. 
Once these datasets have been selected and stored in an array called {\tt use\_files},
we move on to the evaluation of the ZLP predictions. 

\item {\it Subtraction.}
 At this step the code uses the function {\tt matching()} to
  implement the matching procedure
  described in Sect.~\ref{sec:results_sample}.
  It also automatically selects the
  values of $\Delta E_{\rm I}$ and $\Delta E_{\rm II}$ for the training session.
  If the user aims to extract the bandgap properties
  from the onset of $I_{\rm inel}$, 
  the  {\tt bandgap()} function can be used to
  fit Eq.~\ref{eq:I1} to the onset region.

Here the code loops over the $N_{\rm rep}$ replicas and reads each prediction from the extrapolation data file {\tt predict\_x}.
For each replica {\tt k}, the code creates a datafile containing the original spectra intensities 
({\tt original['x*']} and {\tt original['y*']}), the predicted ZLP for this replica ({\tt prediction y}) 
and the predicted ZLP after matching with each spectrum ({\tt match *}). 
For each replica we subtract the matched spectrum from the original spectrum 
to obtain the desired subtraction: {\tt dif * = original * - match *}. 
This is done for each of the total of the replicas and all these results are stored in the  {\tt total\_replicas} dataframe. 
This file is saved in `Data/results/replica\_files' such that a user
can retrieve them  at any time to calculate the
statistical estimators such as prediction means and uncertainties. 

\item {\it Evaluate the subtracted spectra.}
Here the code creates a {\tt mean\_rep} file that contains
all the median predictions and the upper and lower bounds of the 68\% confidence intervals for 
the predicted ZLP, matched spectra and the subtracted spectra, for each of the original recorded
spectra originally given as an input. 
A graphical representation
of the result is then produced, showing the original spectrum, the matched
ZLP and the ZLP-subtracted spectrum including uncertainty bounds. 

\end{itemize}

We emphasize that the {\tt predictions\_pretrained\_net.ipynb} script is similar to the 
{\tt predictions.ipynb} script, but 
can be executed stand-alone
without the need to train again the neural networks, provided that
the model parameters corresponding to some previous training with
the desired input settings are available.
The item {\bf load result files} is now replaced by {\bf create result files}, 
which can be done by importing the pre-trained nets from the {\tt Models} folder.

\end{document}